\documentclass[journal=jctcce,manuscript=article,layout=twocolumn]{achemso}

\usepackage{enumitem}

\usepackage[english]{babel}
\usepackage[table]{xcolor}% http://ctan.org/pkg/xcolor
\usepackage{adjustbox}
\usepackage{float}     % To force figure placement with [H]
\usepackage{placeins}  % To control float placement

\usepackage{amsmath}
\usepackage{graphicx}
\usepackage{natbib}
\usepackage{multirow} % For multirow cells
\usepackage[colorlinks=true, allcolors=blue]{hyperref}
\usepackage[utf8x]{inputenc}

\title{QuantumBind-RBFE: Accurate Relative Binding Free Energy Calculations Using Neural Network Potentials}
\author{Francesc Saban\'es Zariquiey}
\affiliation{Acellera Labs, C Dr Trueta 183, 08005, Barcelona, Spain}
\author{Stephen E. Farr}
\affiliation{Acellera Labs, C Dr Trueta 183, 08005, Barcelona, Spain}
\author{Stefan Doerr}
\affiliation{Acellera Therapeutics, 38350 Fremont Blvd 203 Fremont CA, 94536 USA}
\author{Gianni De Fabritiis}
\email{g.defabritiis@acellera.com}
\affiliation{Computational Science Laboratory, Universitat Pompeu Fabra, Barcelona Biomedical Research Park (PRBB), C Dr. Aiguader 88, 08003, Barcelona, Spain}
\altaffiliation{Acellera Therapeutics, 38350 Fremont Blvd 203 Fremont CA, 94536 USA}
\altaffiliation{Instituci\'o Catalana de Recerca i Estudis Avan\c{c}ats (ICREA), Passeig Lluis Companys 23, 08010 Barcelona, Spain}

\begin{document}
\begin{abstract}
Accurate prediction of protein-ligand binding affinities is crucial in drug discovery, particularly during hit-to-lead and lead optimization phases, however, limitations in ligand force fields continue to impact prediction accuracy. In this work, we validate relative binding free energy (RBFE) accuracy using neural network potentials (NNPs) for the ligands. We utilize a novel NNP model, AceFF 1.0, based on the TensorNet architecture for small molecules that broadens the applicability to diverse drug-like compounds, including all important chemical elements and supporting charged molecules. Using established benchmarks, we show overall improved accuracy and correlation in binding affinity predictions compared with GAFF2 for molecular mechanics and ANI2-x for NNPs. Slightly less accuracy but comparable correlations with OPLS4. We also show that we can run the NNP simulations at 2 fs timestep, at least two times larger than previous NNP models, providing significant speed gains. The results show promise for further evolutions of free energy calculations using NNPs while demonstrating its practical use already with the current generation. The code and NNP model are publicly available for research use. 
\end{abstract}

\section{Introduction}
Accurate prediction of protein-ligand binding affinities is crucial in drug discovery, particularly in the hit-to-lead and lead optimization stages, where a congeneric series of ligands must be screened efficiently. Among the methods employed, alchemical free energy calculations\cite{jorgensen2004many,abel2017advancing,armacost2020novel,york2023modern,qian2024alchemical}, including relative binding free energy (RBFE) methods, have gained prominence due to their capacity to deliver reliable estimates of binding affinities across a wide range of compounds. Such techniques are widely used in both academia and the pharmaceutical industry.
However, the accuracy of RBFE calculations is hampered by several factors such as convergence of the simulations, omission of relevant chemical effects (e.g. tautomerizations and/or protonation), poor selection of pair edges, poor pose selection and/or the inaccuracy of the ligand force field. In this work, we put a special focus on this last point. 

Accurately capturing molecular interactions across the diverse landscape of drug-like compounds is a challenge for traditional molecular mechanics (MM) force fields like GAFF\cite{wang2004development,wang2006automatic}, CGenFF\cite{vanommeslaeghe2010charmm,vanommeslaeghe2012automation}, and OpenFF\cite{qiu2021development,boothroyd2023development}. These force fields can struggle with rare chemical groups and cannot account for key energetic factors, such as polarization and quantum effects. As a result, they can introduce inaccuracies in binding free energy predictions, especially when handling complex ligand chemistry or significant conformational flexibility. 

Recently, an alternative approach has emerged based on neural network potentials (NNPs) to use fast neural network function approximation of the quantum mechanic's energy surface\cite{duval2023hitchhiker}.  Based on these architectures, a few methods have been made available which we will consider in this work.
MACE-based models\cite{kovacs2023mace} are very accurate but computationally expensive. This is not a problem in material science where the alternative is very expensive quantum chemistry methods. For biology-oriented applications, ANI-models\cite{ANI2x} are much faster but at the expenses of accuracy, still providing a very good trade-off. AIMNet2\cite{anstine2024aimnet2} extends ANI with a different architecture to more elements and molecules, still adding on computational costs. The equivariant transform (ET) architecture\cite{tholke2022torchmd} and TensorNet\cite{TensorNet} aimed to provide state-of-the-art accuracy without sacrificing computational efficiency, being taylored specifically for drug discovery. These two are integrated in the TorchMD-Net software framework \cite{doerr2021torchmd,tholke2022torchmd,pelaez2024torchmd} and a new NNP model has been made available, AceFF\cite{aceff_huggingface}. 
 Molecular dynamics (MD) simulations based on NNPs promise accurate simulations at reasonable costs compared with density functional theory (DFT) methods, but do represent an increase in computational costs compared to traditional molecular mechanics, yet their future is very promising \cite{nnpreviewgianni}.
Machine-learned potentials employ relatively short cutoff distances (typically around 5 Å) for efficiency. However, message-passing architectures—such as those based on TensorNet—effectively extend the receptive field beyond the nominal cutoff.\cite{TensorNet} For example, a single interaction layer with a 5 Å cutoff can yield an effective range of approximately 10 Å. Moreover, in our NNP/MM setup, only the ligand’s intramolecular interactions are modeled by the NNP.  Long-range electrostatic is taken care of by the standard classical potential at this time.

%which is generally sufficient to capture the key short- and intermediate-range intramolecular interactions. In contrast, alternative models like AIMNet2\cite{anstine2024aimnet2} explicitly incorporate long-range electrostatic interactions by predicting partial charges. These differing strategies illustrate the balance between computational efficiency and the need to accurately model both short- and long-range effects.

Recent studies by Karwounopoulos et al.\cite{karwounopoulos2025evaluation} have explored an alternative approach in which torsional profiles in classical force fields are fitted via machine learning as a means to approximate the benefits of using an NNP with mechanical embedding. Their results indicate that this end‐state corrected MM/ML method does not offer significant improvements over standard MM in capturing the energetic contributions of ligand strain. In contrast, NNP/MM approach directly employs a neural network potential to model the complete intramolecular energy surface of the ligand, providing a more comprehensive treatment of ligand strain and other subtle energetic effects.

Recently, we explored the integration of NNPs into RBFE calculations with promising results.\cite{sabanes2024enhancing} Specifically, we used  ANI-2x\cite{ANI2x} with good results.  However, ANI-2x cannot handle charged species and has a restricted range of supported atom elements. These limitations restrict the applicability across the diverse chemical space encountered in drug discovery. Recently, alternative approaches have emerged that address these shortcomings by integrating quantum-mechanical insights into the potential. Notably, the QDpi models\cite{zeng2023qdpi,tao2024amber}, use a fast QM/MM-MLP framework to model both intra- and intermolecular interactions, including challenging cases such as tautomers and different protonation states. These models have demonstrated improved performance in reproducing properties like proton affinities. Furthermore, Crha et al. extended their QM/MM framework by integrating a machine‐learned potential to predict both the QM region and its fully polarized buffer’s energies, thereby enabling direct alchemical free‐energy perturbations as demonstrated in the methanol-to-methane transformation in water.\cite{crha2025alchemical}
In this work, we extend our previous work\cite{sabanes2024enhancing} by testing QuantumBind-RBFE, our NNP/MM approach\cite{galvelis2023nnp} for RBFE calculations, using AceFF. The AceFF 1.0\cite{aceff_huggingface} model supports a broad range of atom elements, including charged molecules, and is tailored for accurate and efficient RBFE calculations across diverse molecular systems. 
To validate the performance of RBFE calculations using NNPs, we conducted a comprehensive benchmarking study using established datasets, including both charged and neutral ligands. Our results show that we can improve the accuracy of RBFE calculations compared to traditional force fields and our previous test with  ANI-2x, reaching state-of-the-art correlations. 

\section{Methods}
\subsection{RBFE calculation within the NNP/MM scheme}

\begin{figure}[h!]
\centering
\includegraphics[width=\columnwidth]{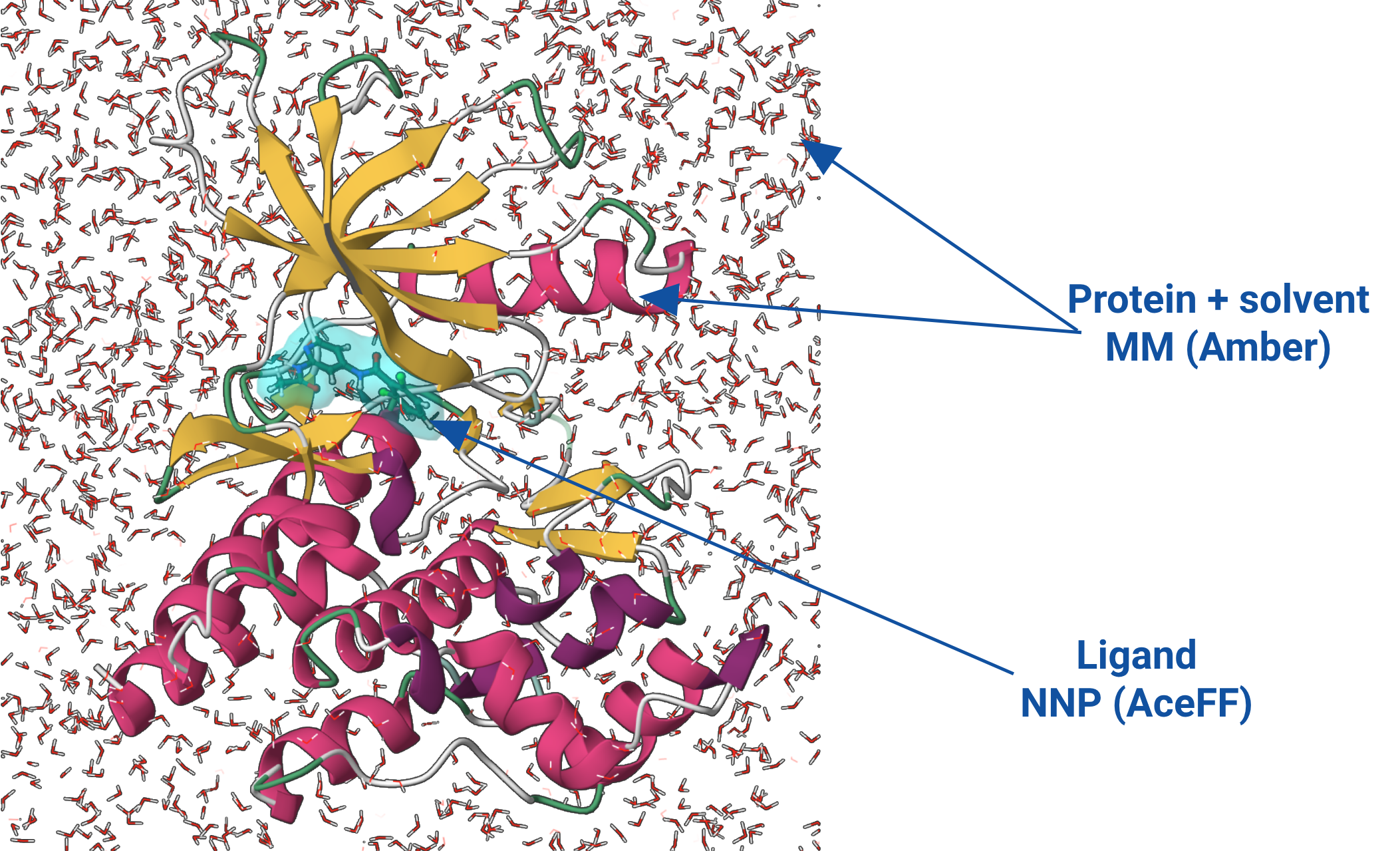}
\caption{Description of the NNP/MM scheme. While the ligand is simulated with a neural network potential (NNP), the rest of the system is treated with classical molecular mechanics (MM)}
\label{fig:NNP_MM}
\end{figure}

Calculations were performed using an NNP/MM approach, which combines neural network potentials (NNP) for high-accuracy modeling of ligand interactions (e.g., capturing internal strain) and molecular mechanics (MM) for the remainder of the system, including all ligand-protein and ligand-solvent interactions.\cite{galvelis2023nnp} This hybrid method allows the ligand to be simulated with the more accurate neural network potential, while the surrounding protein environment is treated with classical molecular mechanics for computational efficiency.(Figure \ref{fig:NNP_MM})
In our NNP/MM scheme, the system is divided into NNP and MM regions, akin to the partitioning used in QM/MM simulations. The total potential energy ($V$) is calculated as:
\begin{equation}
V(\vec{r}) = V_{\text{NNP}}(\vec{r}_{\text{NNP}}) + V_{\text{MM}}(\vec{r}_{\text{MM}}) + V_{\text{NNP-MM}}(\vec{r}),
\end{equation}
where $V_{\text{NNP}}$ describes the ligand's intramolecular interactions using the NNP, and $V_{\text{MM}}$ accounts for the classical MM contributions of the protein and solvent. The coupling term $V_{\text{NNP-MM}}$, representing the nonbonded interactions (electrostatic and van der Waals) between the ligand and its environment is computed entirely using MM. This mechanical embedding scheme ensures that while the ligand–protein interactions are modeled with established MM terms, the ligand’s internal strain—which plays a critical role in binding free energy predictions—is captured at a higher level of theory via the NNP. Although “range corrected” neural network models that incorporate short-range interactions with the MM environment have shown promise in other contexts,\cite{zeng2021development} our current implementation uses a purely mechanical embedding approach.\cite{galvelis2023nnp}

We utilized the Alchemical Transfer Method (ATM) to perform the RBFE calculations.\cite{azimi2022relative} This methodology has been previously validated across multiple benchmarks,\cite{SabanesATM,ATMPsivant} demonstrating its reliability for accurate binding free energy predictions. Notably, ATM was also employed in our recent work validating its use with NNPs, specifically ANI2x.\cite{sabanes2024enhancing} In this study, we extend the previous work substantially.

\subsection{Dataset selection}
We selected a widely recognized public benchmark dataset that serves as a reference for RBFE studies. The BACE, CDK2, JNK1, MCL1, P38, THROMBIN, and TYK2 targets are part of Wang et al.'s dataset, commonly known as the "JACS dataset" or "Schrödinger dataset" \cite{wang2015accurate}. We had to omit PTP1B from the evaluation study since all ligands in the series have charge -2 and AceFF, for now, can only compute ligands with charges -1,0 and +1.
%Additionally, the CDK8, CMET, EG5, HIF2A, PFKFB3, SHP2, SYK, and TNKS2 targets were sourced from the benchmark study by Schindler et al. \cite{schindler2020large}, referred to as the "Merck dataset."
Compared to our previous work,  we can now perform the benchmark in completion. This dataset provides a total of 7 protein targets, 179 ligands, and 280 edges. Comparisons will be made with GAFF2 and FEP+ with the OPLS4 force field\cite{lu2021opls4}.

\subsection{NNP model}

\begin{figure}
\centering
\includegraphics[width=0.5\textwidth]{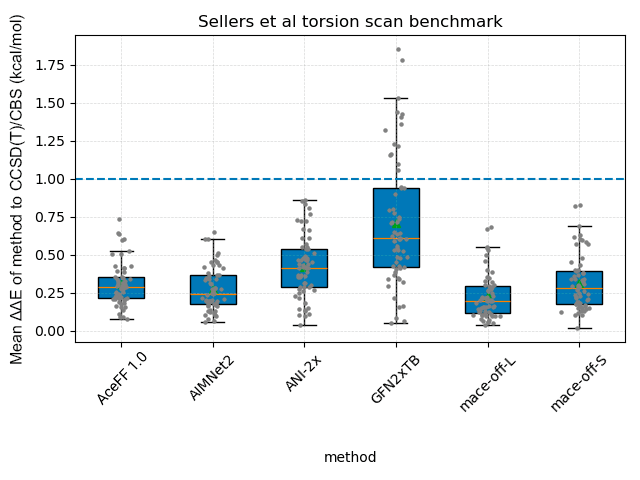}
\caption{Seller's torsion scan benchmark, comparing torsion accuracy for a series of potentials against a CCSD(T)/CBS baseline.}
\label{fig:torsion_scan}
\end{figure}

AceFF 1.0 is the first version of a new family of potentials \cite{aceff_huggingface}. It uses TensorNet 1-layer \cite{TensorNet} trained on Acellera\cite{acellera_website}'s internal proprietary dataset of forces and energies computed the wB97M-V/def2-tzvppd level of theory. This work is the first test of this model for accuracy in RBFE calculations. 
A basic evaluation of the model on the Seller's torsion scan \cite{sellers2017comparison} benchmark shows promising performance  (Figure \ref{fig:torsion_scan}). For all molecules in the test, we performed the full torsion scans, minimizing the geometry with constrained torsion angles using the forces and energy from each calculation method. For comparison, we did the same procedure with ANI-2x\cite{ANI2x} from TorchANI\cite{gao2020torchani}, xTB\cite{bannwarth2019gfn2}, AIMNet2\cite{anstine2024aimnet2}, and MACE-OFF\cite{kovacs2023mace}. We used GeomeTRIC\cite{wang2016geometry} for the constrained optimizations.  The MAE between the curves for each molecule is shown in Figure \ref{fig:torsion_scan}. The AceFF model is among the best performing and comparable to MACE but at a much faster computational speed. AceFF is using a faster TensorNet 1-layer model, instead of the 2-layer model generally used in \cite{TensorNet} for maximum accuracy, which has a speed of 80ns/day for a small molecule\cite{pelaez2024torchmd} (timestep 1fs) compared to MACE-OFF23-Small with 7.5ns/day\cite{kovacs2023mace}.

\subsection{QuantumBind-RBFE Calculation details}

The workflow in this project follows the same procedure as in our previous work\cite{sabanes2024enhancing}. Ligands were parameterized with GAFF2\cite{wang2004development,wang2006automatic} with RESP charges, which are calculated with the OpenFF-Recharge package.\cite{openff_recharge} The ligand topologies were generated using the \textit{parameterize}\cite{galvelis2019scalable} tool. Starting from the protein and prepared ligand structures the protein-ligand complex systems are prepared with HTMD\cite{doerr2016htmd}. Following the protocol from previous works\cite{azimi2022relative,SabanesATM,sabanes2024enhancing} one of the ligands is displaced outside of the binding site via a displacement vector. We ensure that the ligand stays more than 15 \r{A} away from the protein. Next the system is solvated with TIP3P waters with a padding of 10 \r{A} and ions Na\textsuperscript{+} and Cl\textsuperscript{-} are added to neutralize the system at a concentration of 0.15 M. By placing both ligands in the same simulation box with a dual topology approach\cite{gapsys2015calculation}, we avoid issues related to charge-change modifications that can occur when RBFE calculations are performed in separate simulations. Finally, atom indexes for ligand alignment were performed, selecting three reference atoms for each ligand. For more information on the selection of these atoms please check previous publications.\cite{azimi2022relative,SabanesATM}.

The energy minimization, thermalization, and equilibration steps followed the procedures described in our previous work.\cite{SabanesATM}. RBFE simulations were run in triplicate for each edge for an ensemble of 70 ns per replica, a similar amount of simulation time to our previous\cite{sabanes2024enhancing} and similar\cite{hahn2024current} studies.  We used the Amber ff14SB parameters\cite{zou2019blinded,maier2015ff14sb} as well as the TIP3P water model. In all simulations, bonds involving hydrogen atoms were constrained, which mitigates high-frequency vibrational motions and enables the use of larger timesteps. However, we also performed NVE simulations of systems without any constraints and verified that they can similarly conserve the energy as classical MD at the equivalent timestep. Details on the alchemical schedule can be found in Table \ref{tab:params_table}. Classical RBFE simulations were run at a $4$ fs timestep while the NNP/MM runs were run at $2$ fs timestep with a Langevin thermostat for both cases. Previously, we already evaluated how the accuracy difference between 1fs and 4fs runs is minimal for MM runs.\cite{sabanes2024enhancing} In the results section we discuss the reasoning behind selecting such a timestep for the NNP/MM runs.
All calculations were run on GPUGRID.net\cite{gpugrid}. The parallel replica exchange molecular dynamics simulations were conducted using QuantumBind-RBFE our implementation of ATM, HTMD\cite{doerr2016htmd} and the ATMForce potential from OpenMM 8.1.1\cite{eastman2023openmm}.

\subsection{Analysis}
Binding free energies and their corresponding uncertainties from the perturbation energy samples were computed using the Unbinned Weighted Histogram Analysis Method (UWHAM).\cite{tan2012theory} The resulting relative binding free energies ($\Delta \Delta G$) were averaged over the three repeats and used as the predicted value, and the standard deviation across the repeats was used as an error estimate. Absolute $\Delta G$ values were computed via a maximum likelihood estimator with cinnabar for the obtained $\Delta \Delta G$ values.\cite{cinnabar}
The accuracy of the methods was evaluated with several metrics, both from an error and ranking perspective. Relative and absolute binding free energies were compared to experimental measurements of mean absolute error (MAE), root mean square error (RMSE), and Kendall Tau correlation coefficient. Additionally, we evaluated the methods’ ability to prioritize the most active compounds in each dataset by analyzing top compound identification. Specifically, we examined whether the methods correctly identified the top 5 ligands and the top 30\% of ligands ranked by experimental binding affinity. For the top 5 metrics, we assessed how many of the five most active compounds were included in the predicted top 5. For the top 30\% metric, we determined the percentage of ligands within the experimental top 30\% that were also present in the predicted top 30\%. 
All results are reported with 95\% confidence intervals, calculated using 1000 bootstrap samples. To ensure a fair comparison and minimize discrepancies that can appear from differences in algorithm implementation, we recalculated the metrics and converted $\Delta G$ values for OPLS4 with FEP+ using our analysis scripts. These calculations were performed by incorporating the $\Delta\Delta G$ values prior to the cycle closure correction into the algorithm. The initial structures and the results of all RBFE simulations are available on GitHub (refer to the Data Availability section for further details).

\section{Results}

\subsection{QuantumBind-RBFE calculations}
\begin{figure*}
\centering
\includegraphics[width=1\linewidth]{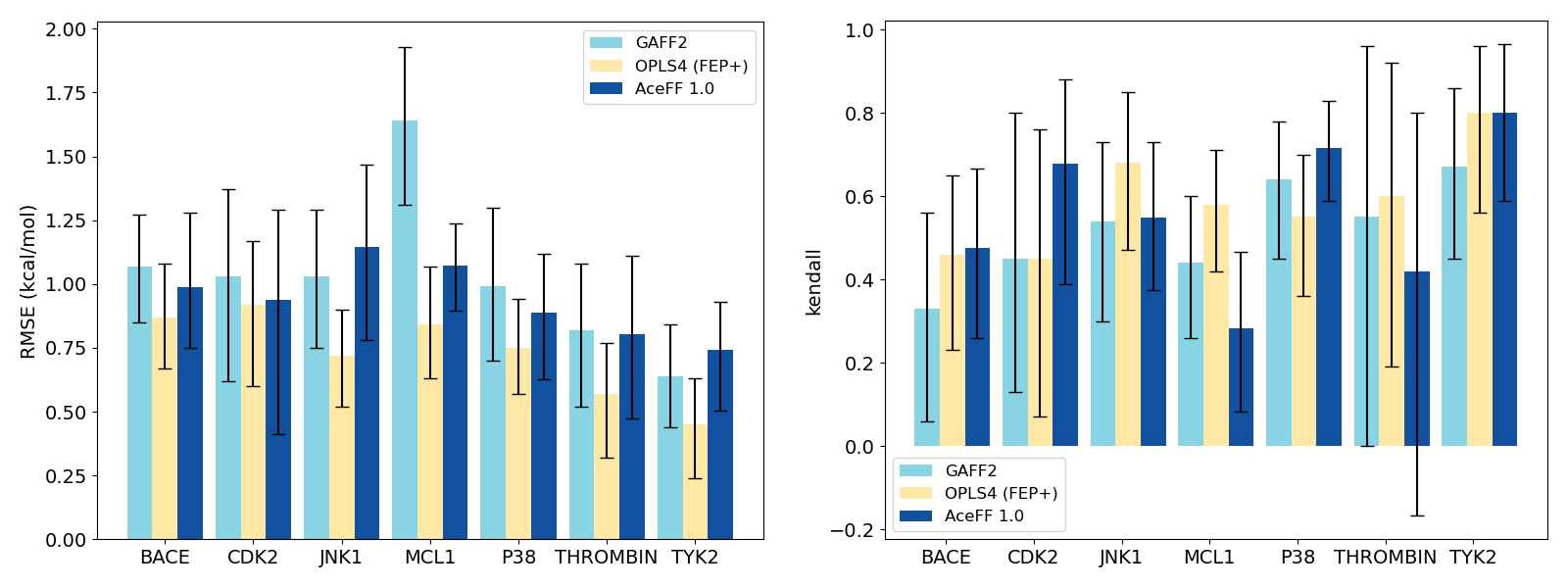}
\caption{(Left) Root Mean Squared Error (RMSE) and (right) Kendall tau correlation for the $\Delta G$s of each protein-ligand system calculated in combination with different approaches:  GAFF2 (teal), reported estimates using FEP+ with the OPLS4 forcefield(yellow) and AceFF 1.0 (blue).}
\label{fig:rmse_kendall}
\end{figure*}

\begin{figure*}
\centering
\includegraphics[width=\linewidth]{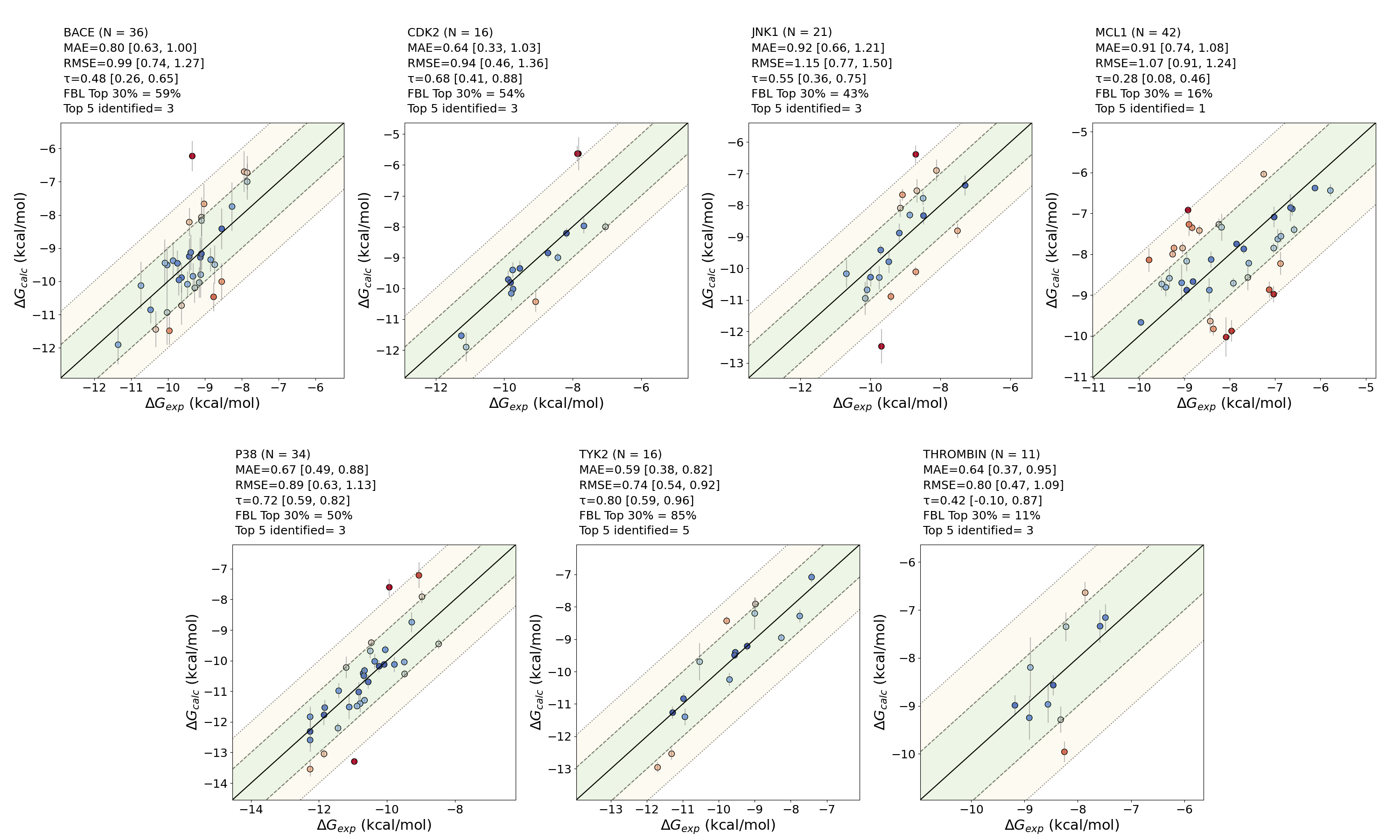}
\caption{Scatterplots of predicted $\Delta G$ values for each evaluated system using AceFF 1.0. The green and yellow shaded areas represent absolute error thresholds of 1 kcal/mol and 2 kcal/mol, respectively. Additional metrics, including mean absolute error (MAE), root mean square error (RMSE), Kendall tau correlation ($\tau$), and top 30\% and top 5 compound identification accuracy, are also displayed. 95\% confidence interval values for the relevant metrics are shown in brackets.}
\label{fig:scatter_DG}
\end{figure*}

%General discussion what works best/worse
%, ranking with kendall but also new metrics
%NNP to MM changes
%    outliers that went well
%    things to worked ok to bad?
Figure \ref{fig:rmse_kendall} and Tables \ref{tab:stats_table},\ref{tab:MAE_pct_table} presents the RMSE, MAE, and Kendall tau correlation values for the evaluated targets, comparing GAFF2, OPLS4, and AceFF 1.0. These metrics provide insight into both the accuracy (via RMSE and MAE) and ranking performance (via Kendall tau) of each method across a diverse set of protein-ligand systems. Scatter plots related to the AceFF 1.0 runs can be found in Figure \ref{fig:scatter_DG}. $\Delta\Delta G$ values for the aforementioned systems and methods are included in the supporting information (Table \ref{tab:stats_ddg_table} and Figure \ref{fig:scatter_ddG_AF}. Since this work uses a different charge model (RESP) compared to our previous publications (AM1BCC), we compared the results obtained here with those from earlier studies (Tables \ref{tab:stats_charges_dG}-\ref{tab:stats_charges_ddG}).
 For most targets, the results are very similar between the two charge models. Notably, in certain cases, the improvement with RESP charges is considerable. While the limitations and possible improvements of this approach are discussed in later sections, the overall improvement observed with RESP charges motivated their use for the QuantumBind-RBFE calculations presented in this work.

QuantumBind-RBFE demonstrates improved performance compared to GAFF2 across most targets. When considering all computed data points, AceFF 1.0 achieves better overall metrics, with a notable decrease in RMSE from 1.17 kcal/mol for GAFF2 to 0.99 kcal/mol for AceFF, and a reduction in MAE from 0.90 to 0.79 kcal/mol. The Kendall tau correlation is slightly better for AceFF 1.0 (0.59 vs. 0.55 for GAFF2), though the difference is relatively modest. Importantly, AceFF consistently achieves a mean absolute error (MAE) below 1.0 kcal/mol for all targets, underscoring its robust accuracy across diverse systems.

%Focusing on individual targets, the RMSE differences between the two methods are minor for BACE, JNK1, P38, and TYK2, ranging from 0.08 to 0.16. This indicates comparable or nearly identical accuracy for these systems. Similarly, MAE values for these targets are also very close between AceFF and GAFF2. The most substantial improvement in error is observed for MCL1, where AceFF 1.0 reduces the RMSE significantly from 1.64 with GAFF2 to 1.07. Likewise, the MAE for MCL1 decreases from 1.32 kcal/mol with GAFF2 to 0.91 kcal/mol with AceFF, further demonstrating AceFF’s advantage in this challenging system.

Despite the similar accuracy in terms of RMSE and MAE across many targets, AceFF 1.0 consistently outperforms GAFF2 in ranking capabilities. Kendall tau correlations are higher for all targets except MCL1 and THROMBIN. The lower correlation for THROMBIN can be attributed to the small size of the dataset (N = 11), where even a single outlier can disproportionately affect the correlation metric, as observed in previous studies.\cite{gapsys2020large} For MCL1, the reduced Kendall tau (0.28 for AceFF vs. 0.44 for GAFF2) is primarily due to the overprediction of binding affinities for five ligands in the series, which significantly impacts ranking performance. 

We analyzed trajectories from both AceFF 1.0 and GAFF2 runs and found that the ligands explore similar conformational spaces in both cases. This indicates that the improved RBFE predictions with AceFF 1.0 mainly come from a better treatment of the ligand’s internal energetics,particularly internal strain, rather than from differences in overall conformational sampling.

AceFF 1.0 demonstrates competitive performance compared to the reported results with OPLS4,\cite{lu2021opls4} though the latter generally achieves lower RMSE and MAE values across most targets. When considering all computed data points, OPLS4 outperforms AceFF 1.0 with a lower RMSE (0.78 vs. 0.99 kcal/mol) and MAE (0.61 vs. 0.79 kcal/mol). Kendall tau correlations are also slightly higher for OPLS4 (0.66 vs. 0.59 for AceFF), indicating better overall ranking accuracy.

%On individual targets, OPLS4 consistently achieves the lowest RMSE and MAE values, particularly on JNK1, MCL1, and TYK2. For instance, on JNK1, OPLS4 achieves an RMSE of 0.72 compared to 1.15 for AceFF and an MAE of 0.59 versus 0.92. Similarly, for MCL1, OPLS4 records an RMSE and MAE of 0.84 and 0.65, respectively, significantly lower than AceFF’s values of 1.07 and 0.91. On TYK2, OPLS4 achieves an RMSE of 0.45 and an MAE of 0.34, outperforming AceFF’s RMSE of 0.74 and MAE of 0.59.

%However, AceFF 1.0 achieves comparable performance to OPLS4 on certain targets. For instance, on BACE, AceFF achieves a Kendall tau of 0.48, similar to OPLS4’s 0.46, despite OPLS4 having a slightly lower RMSE (0.87 vs. 0.99). On P38, AceFF 1.0 achieves a Kendall tau of 0.72, outperforming OPLS4’s 0.55, although OPLS4 achieves a lower RMSE (0.75 vs. 0.89). THROMBIN presents a unique case due to the small size of the dataset (N = 11). Here, OPLS4 achieves the best overall metrics with an RMSE of 0.57 and a Kendall tau of 0.60. While AceFF remains competitive in RMSE (0.80), its lower Kendall tau (0.42) may result from the sensitivity of ranking metrics to outliers in smaller datasets.

Overall,  OPLS4 show still superior RMSE for most targets, while comparable ranking performance to AceFF 1.0 where some targets are better ranked by OPLS4, some by AceFF 1.0. It is worth noting that the methodologies for RBFE calculations differ between QuantumBind and FEP+, including distinct protocols and molecular dynamics engines. These differences can introduce additional variability in the results. The observed differences are also usually small and often within the margin of error, emphasizing the promise of NNPs as a viable alternative for RBFE calculations.

Finally, we highlight the significant improvement AceFF 1.0 provides over GAFF2 when using the same protocol and molecular dynamics engine. This underscores the robustness of AceFF 1.0 as an advanced tool for RBFE predictions in drug discovery workflows. The parameterization of small molecules with NNPs, even using the NNP/MM scheme, becomes trivial, as it does not require the calculation and fitting of torsion scans, greatly simplifying the process.

%JNK1, crystal structure not optimal. Different orientations could be explored, and recent studies with FEP+ showed that reduce error.

\subsection{Identifying the top compounds in a series}

In benchmark studies, the primary objective is often to achieve the highest accuracy and minimize prediction errors. However, in practical drug discovery settings, such as hit-to-lead or lead optimization campaigns, the primary goal is to identify the most active compounds from a pool of candidates, even if their predicted affinities are somewhat overestimated. This prioritization aids in streamlining the selection process for synthesis or purchasing decisions.

To evaluate the practical utility in such scenarios, we assessed its ability to rank ligands correctly across different levels of priority. Specifically, we examined the top 5 ligands and the top 30\% of ligands. The top 5 was measured by the Overlap Coefficient, which quantifies the proportion of correctly ranked ligands among the true top 5. The top 30\% was assessed using the Fraction of Best Ligands (FBL) metric, which evaluates the proportion of ligands in the predicted top 30\% that align with the true top 30\%. This dual assessment allows us to determine not only the accuracy of identifying the very top-ranked ligands but also how well the method performs across a broader range of predicted positions.

\begin{figure*}
\centering
\includegraphics[width=\linewidth]{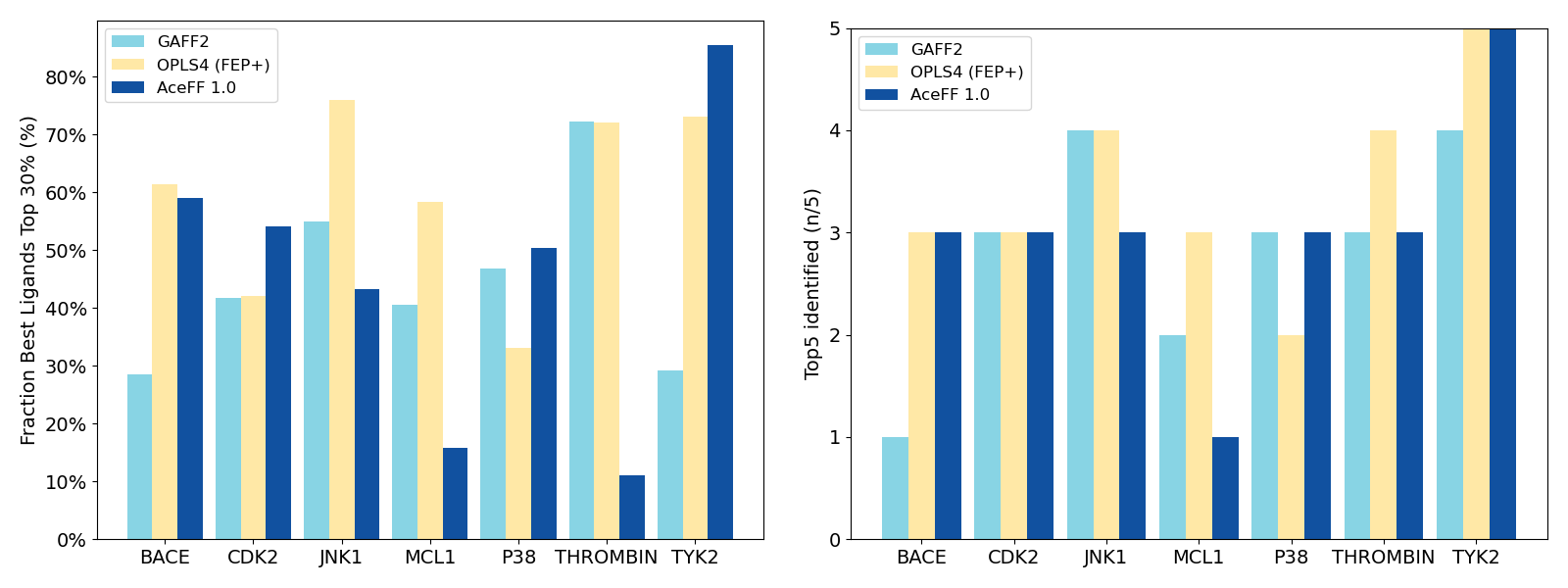}
\caption{Identification of top compounds across datasets. The left plot shows the accuracy in identifying compounds within the top 30\%, while the right plot focuses on the top 5 compounds in each series. Results are compared across the evaluated methods: AceFF 1.0, GAFF2, and OPLS4 with FEP+.}
\label{fig:top_pct}
\end{figure*}

Results (Figure \ref{fig:top_pct} and Table \ref{tab:topid_table}) indicate that AceFF 1.0 generally performs as well as or better than the other methods in identifying the most potent compounds. For example, in datasets such as BACE, CDK2, P38, and TYK2, AceFF 1.0 successfully identified at least 50\% of the top 30\% compounds (59\%, 52\%, 50\%, and 84\%, respectively), outperforming GAFF2 runs in all cases. However, the identification of top compounds by AceFF 1.0 is less effective in some datasets, such as JNK1, MCL1, and THROMBIN, where the top 30\% identification is considerably lower. 
We believe that this metric is particularly valuable for larger datasets such as BACE, MCL1, and P38. For these datasets, AceFF 1.0 generally shows improved performance compared to GAFF2, though with variability across targets: BACE and P38 exhibit clear improvements, while MCL1 remains a challenge.
%The choice of the percentage threshold for evaluation (e.g., 10\%, 30\%, or 50\%) may depend on the size of the dataset and the distribution of $\Delta G$ values within the congeneric series. For our study, we selected 30\% as a middle ground due to the variation in dataset sizes and the practicality of this threshold for compound prioritization.

In terms of top 5 identification, the differences between methods were generally small. For MCL1, AceFF 1.0 placed one ligand in the top 5, while GAFF2 and OPLS4 identified two and three, respectively. In most other cases, the same number of top-ranked molecules were identified, with discrepancies of no more than a single ligand when comparing AceFF 1.0 to OPLS4 (Figure \ref{fig:top_pct}). Notably, even in cases where AceFF 1.0 exhibited higher overall error rates compared to OPLS4, its compound prioritization remained comparable. 
%In many instances, AceForce 1.0 even outperformed OPLS4 in identifying the most active compounds, demonstrating its utility for practical drug discovery workflows.

\subsection{Timesteps: 1fs vs 2fs}
%Discussion on why we decided to keep 1fs results
%Discussion on smoothness of potential
One of the key challenges of NNP/MM RBFE calculations is the increased computational cost required to execute these simulations. This is primarily due to two factors: the overhead associated with running the NNP and the limitations on the simulation timestep. While MM calculations can be confidently performed with a 4 fs timestep, NNP/MM runs have traditionally been restricted to 1 fs. Previous attempts to increase the timestep frequently resulted in simulation failures.

\begin{figure}[!ht]
\centering
\includegraphics[width=\columnwidth]{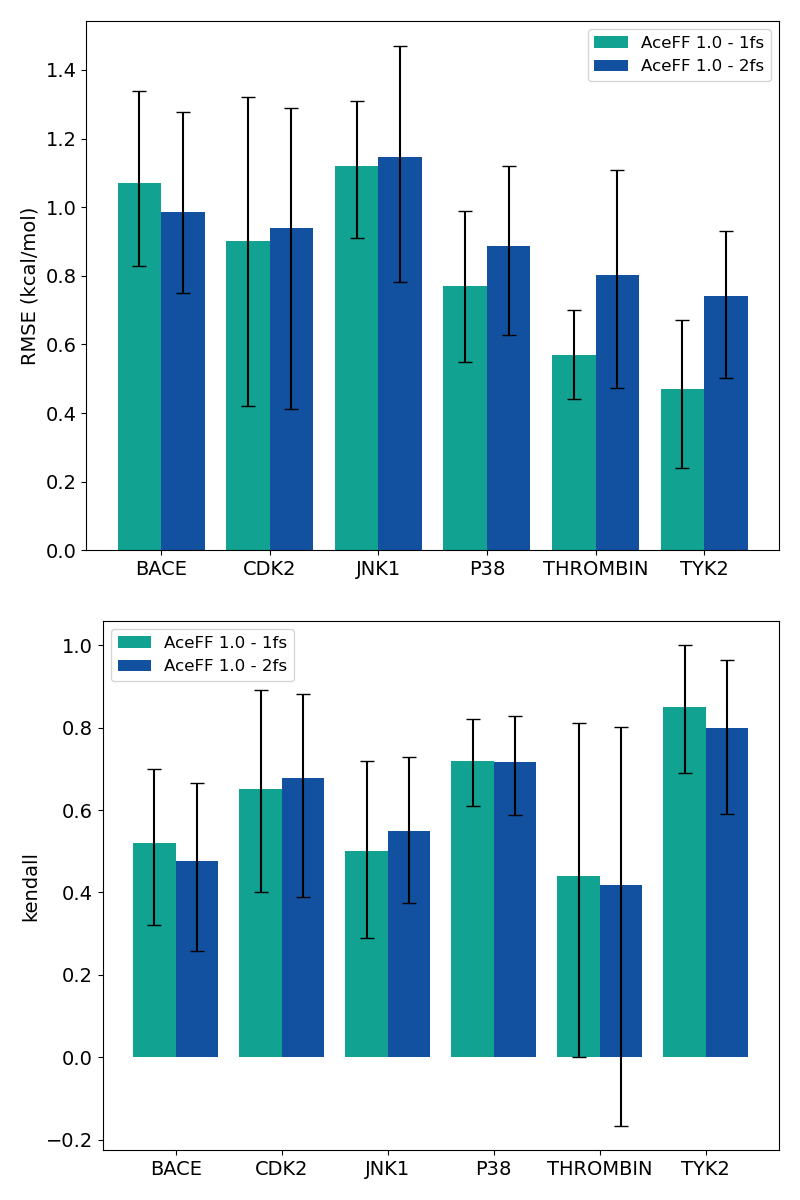}
\caption{Comparison of AceFF 1.0 model accuracy at 1fs and 2fs timesteps. The top panel shows Root Mean Squared Error (RMSE) and the bottom panel shows Kendall tau correlation for $\Delta G$ values across a subset of systems. While runs at both timesteps yield similar performance, the 1fs timestep calculations maintain slightly higher accuracy.}
\label{fig:rmse_kendall_timestep}
\end{figure}

With the AceFF 1.0 model for TorchMD-Net, we observed improved stability at a 2fs timestep, significantly enhancing the computational speed of NNP/MM simulations.(Figure \ref{fig:QB_speeds}) To evaluate whether this larger timestep maintains accuracy, we conducted RBFE calculations on a subset of the JACS dataset at a timestep of both 1 and 2fs. %Later on we'll have Merck and other datasets as well, so we will only do the comparison with JACS
Our findings (Figure \ref{fig:rmse_kendall_timestep} and  Tables \ref{tab:stats_timestep_dG} and \ref{tab:stats_timestep_ddG}) indicate that 2 fs timestep simulations demonstrate comparable accuracy to 1 fs runs for most targets. In these cases, RMSE and Kendall tau correlation values remained nearly identical. The impact on top compound identification is shown in Table \ref{tab:topid_table_timestep}.  Considering the variability between runs, we can conclude that it is possible to run at simulations at 2 fs.

%In BACE, the top 30\% identification dropped from 75\% at 1fs to 59\% at 2fs and in THROMBIN goes from 56\% to 11\% respectively. On the other hand, an improvement is seen in JNK1, where an improvement is observed (32\% to 43\%).  

\subsection{Comparison with ANI-2x}
% Compared with our previous results with ANI-2x
% Selecting only P38 and TYK2 since they are the only targets we could do in full due to ANI-2x limitations.
%Results very similar with AceFF sligthly overperforming in both systems.

\begin{figure}[htb!]
\centering
\includegraphics[width=0.8\columnwidth]{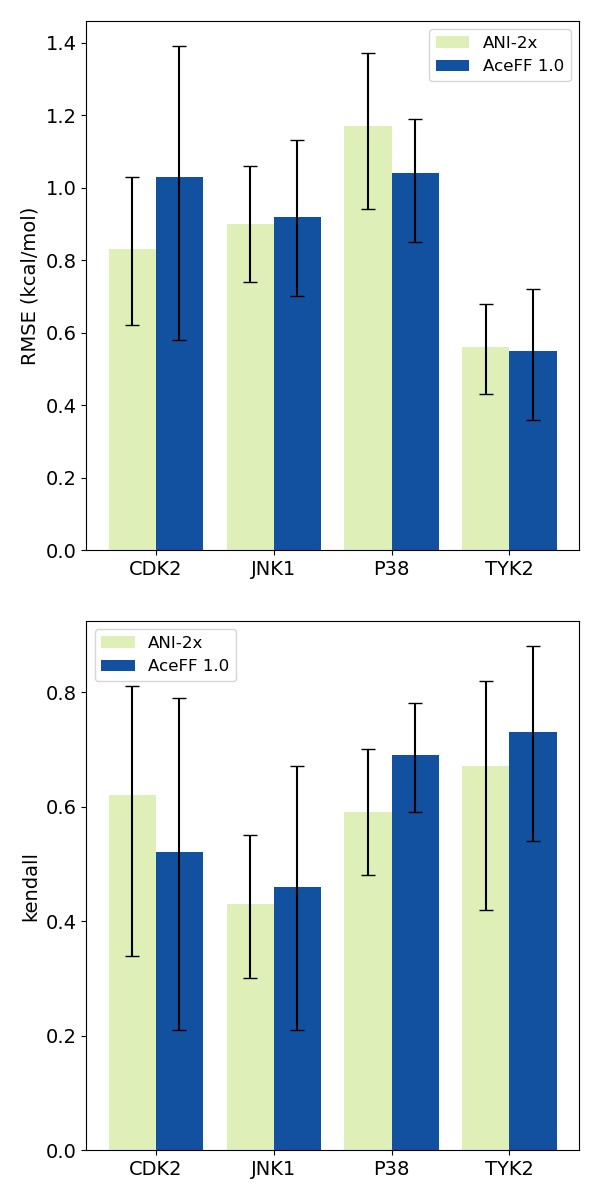}
\caption{Comparison of AceFF and ANI-2x calculations for $\Delta\Delta G$ values.}
\label{fig:ani2x_compare}
\end{figure}

We compare now the performance of AceFF to ANI-2x in terms of accuracy.\cite{sabanes2024enhancing} From the JACS dataset, we computed several edges for CDK2, JNK1, P38, and TYK2. Due to limitations in ANI-2x’s atom type and charge handling, the datasets for JNK1 and CDK2 are not complete, so we compare only the same edges that both methods could calculate. In contrast, the P38 and TYK2 datasets are complete for both models, allowing for a more direct assessment of their performance. For this reason, a comparison between $\Delta G$ values is only done for P38 and TYK2.
RMSE and MAE values for the computed $\Delta\Delta G$ edges are presented in Figure \ref{fig:ani2x_compare}. Additionally, Table \ref{tab:stats_ani} presents a summary of the $\Delta G$ results, while Table \ref{tab:stats_ddg_ani} provides the $\Delta\Delta G$ statistics for all four targets.
For $\Delta\Delta G$ calculations on P38, AceFF achieves a slightly lower RMSE of 1.04 kcal/mol compared to ANI-2x’s 1.17 kcal/mol, alongside a lower MAE (0.84 vs. 0.91 kcal/mol). Additionally, AceFF shows a higher Kendall tau correlation (0.69 vs. 0.59). 
On TYK2, both models deliver similar accuracy, with AceFF showing a slightly lower RMSE (0.55 vs. 0.56 kcal/mol) and a modestly higher Kendall tau (0.73 vs. 0.67). For CDK2, AceFF yields an RMSE of 1.03 kcal/mol, slightly higher than ANI-2x’s 0.83 kcal/mol, and a Kendall tau of 0.46 compared to ANI-2x’s 0.62, indicating that ANI-2x captures the ranking of these particular ligands more effectively. Conversely, on JNK1, AceFF attains an RMSE of 0.92 kcal/mol vs. ANI-2x’s 0.90 kcal/mol, and a higher Kendall tau (0.46 vs. 0.43), suggesting a marginal ranking advantage for AceFF on this target.
For absolute binding free energies ($\Delta G$), AceFF again demonstrates improved overall performance. 
On P38, AceFF achieves an RMSE of 0.77 kcal/mol and a Kendall tau of 0.72, and ANI-2x (RMSE 0.93, Kendall tau 0.58). Similarly, on TYK2, AceFF reaches an RMSE of 0.47 kcal/mol and Kendall tau of 0.85, slightly surpassing ANI-2x’s RMSE of 0.50 and Kendall tau of 0.82. 
Overall, these findings indicate that AceFF  generally perform better across the evaluated systems with the additional benefit that it is not limited to those.

\subsection{Going beyond the limits of the model}

\begin{figure*}[htb!]
\centering
\includegraphics[width=2\columnwidth]{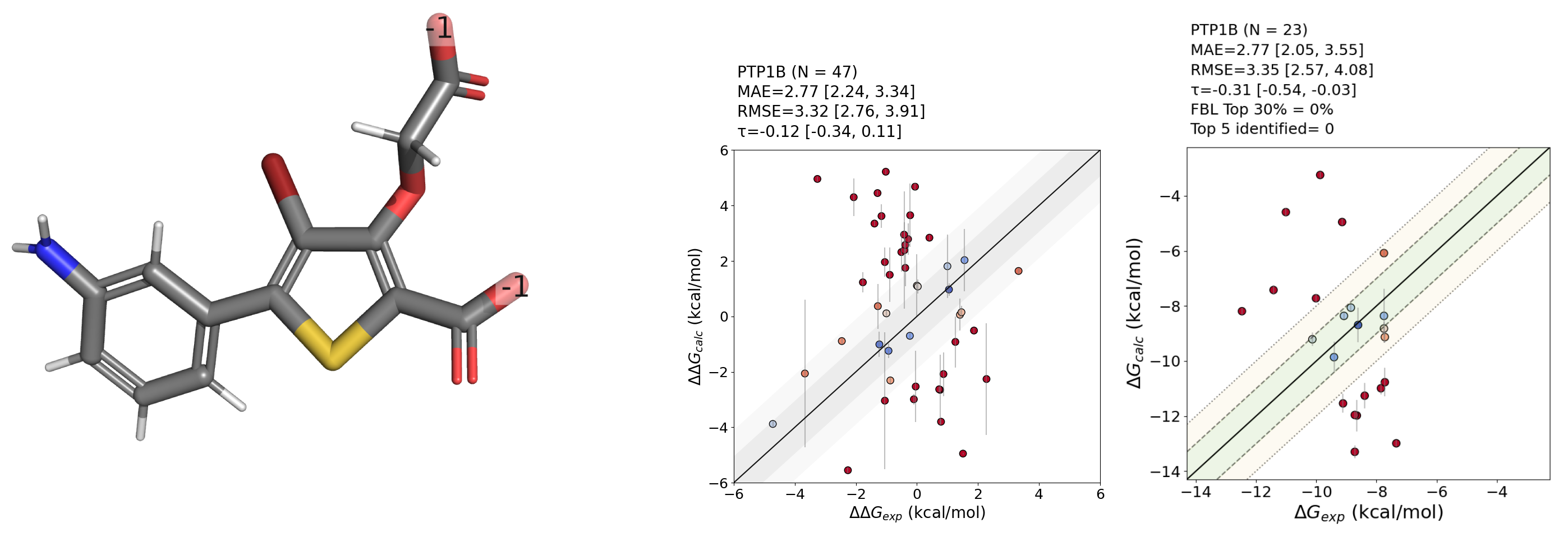}
\caption{Left: Example on one of the ligands of the PTP1B dataset, highlighting the -1 charge at each of the two COO- groups. Right: scatterplot for the relative and absolute binding free energy calculations. Relevant metrics are reported on top of each plot}
\label{fig:PTP1B_example}
\end{figure*}

PTP1B is a protein target included in the JACS dataset, characterized by ligands uniformly carrying a -2 charge. AceFF 1.0, however, has been  trained on molecules with charges limited to -1, 0, and +1, making its applicability for this target troublesome. This limitation of AceFF will be solved in version 1.1. We conducted RBFE calculations for PTP1B to assess the model's performance under such constraints. As anticipated, the predictions exhibited significant errors, with RMSE exceeding 3 kcal/mol for both $\Delta\Delta G$ and $\Delta G$ values and two out of the 49 runs crashing. Furthermore, the correlation performance was notably poor, reflected by a Kendall tau coefficient of -0.31, and the method failed to correctly identify any of the top ligands in the dataset (Figure \ref{fig:PTP1B_example}).  This outcome aligns with expectations, emphasizing the need for further training of the model on an extended range of ligand charges to improve its generalization and predictive capabilities.

\section{Conclusions}
% We delevoped a novel NNP, AceFF that overcomes the limitatons of existing ones, such as extended atom types and support for charged systems.
% We evaluated the accuracy of this NNP, via an RBFE benchmark where we compared with an MM approach and the SOTA option.
% AceFF 1.0 generally has equal or better performance than MM runs, with lower errors, higher correlations and generally does a better job at identifying the top compounds in a series. 
% There are still some targets where the error is higher than expected which could be due to the training dataset of the NNP model.
% Upon comparing to FEP+, the SOTA option, AceFF 1.0 has a comparable or better performance for half of the evaluated systems, but underperforms for the rest.
% AceFF 1.0 has a comparable or even slightly better performance than ANI-2x, the initial NNP we tested in a previous publication.
% We could try another MD engine/RBFE protocol (OpenFE?)
% We could try another protein and ligand forcefield (Open FF)
% AceFF 1.0 is a novel NNP that allows computation of an extended number atom types and charged molecules that others could not
% AceFF 1.0 shows an increase in accuracy compared to MM runs in the same setting, but it still has room to improve to achieve state of the art accuracy. 
%Speed is still a limiting factor. We tested the feasability of increasing the timestep from 1 to 2fs, and although they ran without crashes, we observe a consistend drop in accuracy.

We are presenting our NNP/MM approach to perform RBFE calculations. We are using AceFF 1.0, a novel NNP model that supports a broader range of atom elements and charged molecules. To assess its accuracy, we benchmarked AceFF 1.0 using the JACS dataset\cite{wang2015accurate} (without PTP1B due to charge limitations) and compared it to an MM-based approach as well as the industry state-of-the-art alternative in OPLS4 with FEP+.\cite{lu2021opls4}
AceFF 1.0 demonstrated overall improved  performance to MM-based calculations like GAFF2. These improvements highlight NNP potentials as an accurate alternative to MM approaches for RBFE calculations. 
In comparison to OPLS4, the state-of-the-art approach in the industry, AceFF 1.0 provided competitive or superior correlation performance on half of the evaluated systems but underperformed on the rest. These results indicate that further training of the potential may be needed to consistently achieve SOTA accuracy across all targets. Additionally, when compared to ANI-2x, the initial NNP tested in our previous work,\cite{sabanes2024enhancing} AceFF 1.0 delivered better accuracy overall.

It is important to note, however, that the performance of NNPs is strongly dependent on the diversity of their training datasets. For instance, AceFF 1.0 was trained only on molecules with charges of –1, 0, and +1. Consequently, it does not perform well for systems involving ligands with rare charge states, as observed with the PTP1B dataset. This limitation underscores the need for expanding the training data to encompass a broader chemical space, which we plan to address in future versions of AceFF.

One of the main limitations for NNPs remains computational speed, primarily due to timestep restrictions. To enhance computational efficiency, we demonstrated the feasibility of increasing the timestep from 1fs to 2 fs. This adjustment allows for stable simulations with similar accuracy in most cases. The results suggest that a 2 fs timestep can be used for more efficient drug discovery workflows.

%To explore the full potential of AceFF 1.0, future studies could test it across alternative RBFE protocols and force fields, such as OpenFE and OpenFF respectively, to further understand its compatibility and performance.
Future work will focus on testing and developing new versions of AceFF with an expanded training dataset to enhance predictive accuracy and general applicability. We also aim to enable stable simulations at larger timesteps, potentially 3fs or even 4fs, to reduce the computational cost of NNP/MM calculations to levels comparable with MM-based methods. Additionally, we plan to expand RBFE studies to include other benchmark datasets.

\section{Acknowledgement}
The authors thank the volunteers of GPUGRID.net for donating computing time. This project has received funding from
 the Torres-Quevedo Programme from the Spanish National Agency for Research (PTQ2023-012967 / AEI / 10.13039/501100011033).

\section{Conflict of interests}
All the authors have a potential conflict of interest due to direct interests in Acellera Therapeutics (See affiliations). 
\section{Author Contributions}
F.S.Z wrote the manuscript and performed the RBFE benchmark; S.E.F and S.D built AceFF 1.0 and developed QuantumBind-RBFE; G.D.F designed the project and wrote the manuscript
\section{Associated Content}
The Supporting Information contains:
\begin{itemize}[leftmargin=2pt]
\item  Alchemical Schedule: A detailed table showing the alchemical schedule of the Softplus Alchemical Potential parameters used for the two legs of the alchemical transformations.
\item Free Energy Statistics: Comprehensive tables summarizing the raw 
$\Delta G$ and $\Delta \Delta G$ values, along with RMSE, MAE, and Kendall tau correlation metrics for all evaluated targets across the different methods (GAFF2, OPLS4 with FEP+, and AceFF 1.0).
\item Charge Model Comparisons: Detailed comparisons between RESP and AM1BCC charge models, including performance metrics and top identifier analyses.
\item Timestep comparisons: additional tables comparing AceFF at 1 and 2fs timesteps.
\item Comparative Analyses: additional tables comparing AceFF and ANI-2x calculations for both $\Delta G$ and $\Delta \Delta G$ values
\item Simulation speed benchmarks: a figure illustrating the simulation speed for various modes.
\end{itemize}
\section{Data and software availability}
\begin{itemize}[leftmargin=2pt]

\item AceFF models are available from HuggingFace \url{https://huggingface.co/Acellera/AceFF-1.0}
\item Examples of how to use it are available:
  \begin{itemize}[leftmargin=*, label=--]
\item Run ML potential molecular simulations of a small molecule using ACEMD \url{https://software.acellera.com/acemd/nnp.html}.
\item For a tutorial on running mixed protein-ligand simulations, refer to NNP/MM \url{https://software.acellera.com/acemd/nnpmm.html}.
  \end{itemize}

\item A tutorial on how to run the RBFE calculations, input data and structures are available at \url{https://github.com/Acellera/quantumbind_rbfe}.
\end{itemize}

\clearpage
\onecolumn

\bibliography{ref}

\providecommand{\latin}[1]{#1}
\makeatletter
\providecommand{\doi}
  {\begingroup\let\do\@makeother\dospecials
  \catcode`\{=1 \catcode`\}=2 \doi@aux}
\providecommand{\doi@aux}[1]{\endgroup\texttt{#1}}
\makeatother
\providecommand*\mcitethebibliography{\thebibliography}
\csname @ifundefined\endcsname{endmcitethebibliography}  {\let\endmcitethebibliography\endthebibliography}{}
\begin{mcitethebibliography}{51}
\providecommand*\natexlab[1]{#1}
\providecommand*\mciteSetBstSublistMode[1]{}
\providecommand*\mciteSetBstMaxWidthForm[2]{}
\providecommand*\mciteBstWouldAddEndPuncttrue
  {\def\EndOfBibitem{\unskip.}}
\providecommand*\mciteBstWouldAddEndPunctfalse
  {\let\EndOfBibitem\relax}
\providecommand*\mciteSetBstMidEndSepPunct[3]{}
\providecommand*\mciteSetBstSublistLabelBeginEnd[3]{}
\providecommand*\EndOfBibitem{}
\mciteSetBstSublistMode{f}
\mciteSetBstMaxWidthForm{subitem}{(\alph{mcitesubitemcount})}
\mciteSetBstSublistLabelBeginEnd
  {\mcitemaxwidthsubitemform\space}
  {\relax}
  {\relax}

\bibitem[Jorgensen(2004)]{jorgensen2004many}
Jorgensen,~W.~L. The many roles of computation in drug discovery. \emph{Science} \textbf{2004}, \emph{303}, 1813--1818\relax
\mciteBstWouldAddEndPuncttrue
\mciteSetBstMidEndSepPunct{\mcitedefaultmidpunct}
{\mcitedefaultendpunct}{\mcitedefaultseppunct}\relax
\EndOfBibitem
\bibitem[Abel \latin{et~al.}(2017)Abel, Wang, Harder, Berne, and Friesner]{abel2017advancing}
Abel,~R.; Wang,~L.; Harder,~E.~D.; Berne,~B.; Friesner,~R.~A. Advancing drug discovery through enhanced free energy calculations. \emph{Acc. Chem. Res.} \textbf{2017}, \emph{50}, 1625--1632\relax
\mciteBstWouldAddEndPuncttrue
\mciteSetBstMidEndSepPunct{\mcitedefaultmidpunct}
{\mcitedefaultendpunct}{\mcitedefaultseppunct}\relax
\EndOfBibitem
\bibitem[Armacost \latin{et~al.}(2020)Armacost, Riniker, and Cournia]{armacost2020novel}
Armacost,~K.~A.; Riniker,~S.; Cournia,~Z. Novel Directions in Free Energy Methods and Applications. \emph{J. Chem. Inf. Model.} \textbf{2020}, \emph{60}, 1--5, PMID: 31983210\relax
\mciteBstWouldAddEndPuncttrue
\mciteSetBstMidEndSepPunct{\mcitedefaultmidpunct}
{\mcitedefaultendpunct}{\mcitedefaultseppunct}\relax
\EndOfBibitem
\bibitem[York(2023)]{york2023modern}
York,~D.~M. Modern alchemical free energy methods for drug discovery explained. \emph{ACS Physical Chemistry Au} \textbf{2023}, \emph{3}, 478--491\relax
\mciteBstWouldAddEndPuncttrue
\mciteSetBstMidEndSepPunct{\mcitedefaultmidpunct}
{\mcitedefaultendpunct}{\mcitedefaultseppunct}\relax
\EndOfBibitem
\bibitem[Qian \latin{et~al.}(2024)Qian, Xue, Xu, and Huang]{qian2024alchemical}
Qian,~R.; Xue,~J.; Xu,~Y.; Huang,~J. Alchemical Transformations and Beyond: Recent Advances and Real-World Applications of Free Energy Calculations in Drug Discovery. \emph{Journal of Chemical Information and Modeling} \textbf{2024}, \emph{64}, 7214--7237\relax
\mciteBstWouldAddEndPuncttrue
\mciteSetBstMidEndSepPunct{\mcitedefaultmidpunct}
{\mcitedefaultendpunct}{\mcitedefaultseppunct}\relax
\EndOfBibitem
\bibitem[Wang \latin{et~al.}(2004)Wang, Wolf, Caldwell, Kollman, and Case]{wang2004development}
Wang,~J.; Wolf,~R.~M.; Caldwell,~J.~W.; Kollman,~P.~A.; Case,~D.~A. Development and testing of a general amber force field. \emph{Journal of computational chemistry} \textbf{2004}, \emph{25}, 1157--1174\relax
\mciteBstWouldAddEndPuncttrue
\mciteSetBstMidEndSepPunct{\mcitedefaultmidpunct}
{\mcitedefaultendpunct}{\mcitedefaultseppunct}\relax
\EndOfBibitem
\bibitem[Wang \latin{et~al.}(2006)Wang, Wang, Kollman, and Case]{wang2006automatic}
Wang,~J.; Wang,~W.; Kollman,~P.~A.; Case,~D.~A. Automatic atom type and bond type perception in molecular mechanical calculations. \emph{J. Mol. Graphics Modell.} \textbf{2006}, \emph{25}, 247--260\relax
\mciteBstWouldAddEndPuncttrue
\mciteSetBstMidEndSepPunct{\mcitedefaultmidpunct}
{\mcitedefaultendpunct}{\mcitedefaultseppunct}\relax
\EndOfBibitem
\bibitem[Vanommeslaeghe \latin{et~al.}(2010)Vanommeslaeghe, Hatcher, Acharya, Kundu, Zhong, Shim, Darian, Guvench, Lopes, and Vorobyov]{vanommeslaeghe2010charmm}
Vanommeslaeghe,~K.; Hatcher,~E.; Acharya,~C.; Kundu,~S.; Zhong,~S.; Shim,~J.; Darian,~E.; Guvench,~O.; Lopes,~P.; Vorobyov,~A.,~Igor MacKerell~Jr CHARMM general force field: A force field for drug-like molecules compatible with the CHARMM all-atom additive biological force fields. \emph{Journal of computational chemistry} \textbf{2010}, \emph{31}, 671--690\relax
\mciteBstWouldAddEndPuncttrue
\mciteSetBstMidEndSepPunct{\mcitedefaultmidpunct}
{\mcitedefaultendpunct}{\mcitedefaultseppunct}\relax
\EndOfBibitem
\bibitem[Vanommeslaeghe \latin{et~al.}(2012)Vanommeslaeghe, Raman, and MacKerell~Jr]{vanommeslaeghe2012automation}
Vanommeslaeghe,~K.; Raman,~E.~P.; MacKerell~Jr,~A.~D. Automation of the CHARMM General Force Field (CGenFF) II: assignment of bonded parameters and partial atomic charges. \emph{Journal of chemical information and modeling} \textbf{2012}, \emph{52}, 3155--3168\relax
\mciteBstWouldAddEndPuncttrue
\mciteSetBstMidEndSepPunct{\mcitedefaultmidpunct}
{\mcitedefaultendpunct}{\mcitedefaultseppunct}\relax
\EndOfBibitem
\bibitem[Qiu \latin{et~al.}(2021)Qiu, Smith, Boothroyd, Jang, Hahn, Wagner, Bannan, Gokey, Lim, Stern, \latin{et~al.} others]{qiu2021development}
Qiu,~Y.; Smith,~D.~G.; Boothroyd,~S.; Jang,~H.; Hahn,~D.~F.; Wagner,~J.; Bannan,~C.~C.; Gokey,~T.; Lim,~V.~T.; Stern,~C.~D.; others Development and benchmarking of open force field v1. 0.0—the parsley small-molecule force field. \emph{Journal of chemical theory and computation} \textbf{2021}, \emph{17}, 6262--6280\relax
\mciteBstWouldAddEndPuncttrue
\mciteSetBstMidEndSepPunct{\mcitedefaultmidpunct}
{\mcitedefaultendpunct}{\mcitedefaultseppunct}\relax
\EndOfBibitem
\bibitem[Boothroyd \latin{et~al.}(2023)Boothroyd, Behara, Madin, Hahn, Jang, Gapsys, Wagner, Horton, Dotson, Thompson, \latin{et~al.} others]{boothroyd2023development}
Boothroyd,~S.; Behara,~P.~K.; Madin,~O.~C.; Hahn,~D.~F.; Jang,~H.; Gapsys,~V.; Wagner,~J.~R.; Horton,~J.~T.; Dotson,~D.~L.; Thompson,~M.~W.; others Development and benchmarking of open force field 2.0. 0: the Sage small molecule force field. \emph{Journal of chemical theory and computation} \textbf{2023}, \emph{19}, 3251--3275\relax
\mciteBstWouldAddEndPuncttrue
\mciteSetBstMidEndSepPunct{\mcitedefaultmidpunct}
{\mcitedefaultendpunct}{\mcitedefaultseppunct}\relax
\EndOfBibitem
\bibitem[Duval \latin{et~al.}(2023)Duval, Mathis, Joshi, Schmidt, Miret, Malliaros, Cohen, Lio, Bengio, and Bronstein]{duval2023hitchhiker}
Duval,~A.; Mathis,~S.~V.; Joshi,~C.~K.; Schmidt,~V.; Miret,~S.; Malliaros,~F.~D.; Cohen,~T.; Lio,~P.; Bengio,~Y.; Bronstein,~M. A Hitchhiker's Guide to Geometric GNNs for 3D Atomic Systems. \emph{arXiv preprint arXiv:2312.07511} \textbf{2023}, \relax
\mciteBstWouldAddEndPunctfalse
\mciteSetBstMidEndSepPunct{\mcitedefaultmidpunct}
{}{\mcitedefaultseppunct}\relax
\EndOfBibitem
\bibitem[Kov{\'a}cs \latin{et~al.}(2023)Kov{\'a}cs, Moore, Browning, Batatia, Horton, Kapil, Witt, Magd{\u{a}}u, Cole, and Cs{\'a}nyi]{kovacs2023mace}
Kov{\'a}cs,~D.~P.; Moore,~J.~H.; Browning,~N.~J.; Batatia,~I.; Horton,~J.~T.; Kapil,~V.; Witt,~W.~C.; Magd{\u{a}}u,~I.-B.; Cole,~D.~J.; Cs{\'a}nyi,~G. MACE-OFF23: Transferable machine learning force fields for organic molecules. \emph{arXiv preprint arXiv:2312.15211} \textbf{2023}, \relax
\mciteBstWouldAddEndPunctfalse
\mciteSetBstMidEndSepPunct{\mcitedefaultmidpunct}
{}{\mcitedefaultseppunct}\relax
\EndOfBibitem
\bibitem[Devereux \latin{et~al.}(2020)Devereux, Smith, Huddleston, Barros, Zubatyuk, Isayev, and Roitberg]{ANI2x}
Devereux,~C.; Smith,~J.~S.; Huddleston,~K.~K.; Barros,~K.; Zubatyuk,~R.; Isayev,~O.; Roitberg,~A.~E. Extending the Applicability of the ANI Deep Learning Molecular Potential to Sulfur and Halogens. \emph{Journal of Chemical Theory and Computation} \textbf{2020}, \emph{16}, 4192--4202, PMID: 32543858\relax
\mciteBstWouldAddEndPuncttrue
\mciteSetBstMidEndSepPunct{\mcitedefaultmidpunct}
{\mcitedefaultendpunct}{\mcitedefaultseppunct}\relax
\EndOfBibitem
\bibitem[Anstine \latin{et~al.}(2024)Anstine, Zubatyuk, and Isayev]{anstine2024aimnet2}
Anstine,~D.; Zubatyuk,~R.; Isayev,~O. AIMNet2: a neural network potential to meet your neutral, charged, organic, and elemental-organic needs. \emph{ChemrXiv preprint chemrxiv-2023-296ch-v3} \textbf{2024}, \relax
\mciteBstWouldAddEndPunctfalse
\mciteSetBstMidEndSepPunct{\mcitedefaultmidpunct}
{}{\mcitedefaultseppunct}\relax
\EndOfBibitem
\bibitem[Th{\"o}lke and De~Fabritiis(2022)Th{\"o}lke, and De~Fabritiis]{tholke2022torchmd}
Th{\"o}lke,~P.; De~Fabritiis,~G. Torchmd-net: equivariant transformers for neural network based molecular potentials. \emph{arXiv preprint arXiv:2202.02541} \textbf{2022}, \relax
\mciteBstWouldAddEndPunctfalse
\mciteSetBstMidEndSepPunct{\mcitedefaultmidpunct}
{}{\mcitedefaultseppunct}\relax
\EndOfBibitem
\bibitem[Simeon and De~Fabritiis(2023)Simeon, and De~Fabritiis]{TensorNet}
Simeon,~G.; De~Fabritiis,~G. TensorNet: Cartesian Tensor Representations for Efficient Learning of Molecular Potentials. Advances in Neural Information Processing Systems. 2023; pp 37334--37353\relax
\mciteBstWouldAddEndPuncttrue
\mciteSetBstMidEndSepPunct{\mcitedefaultmidpunct}
{\mcitedefaultendpunct}{\mcitedefaultseppunct}\relax
\EndOfBibitem
\bibitem[Doerr \latin{et~al.}(2021)Doerr, Majewski, P{\'e}rez, Kramer, Clementi, Noe, Giorgino, and De~Fabritiis]{doerr2021torchmd}
Doerr,~S.; Majewski,~M.; P{\'e}rez,~A.; Kramer,~A.; Clementi,~C.; Noe,~F.; Giorgino,~T.; De~Fabritiis,~G. TorchMD: A deep learning framework for molecular simulations. \emph{Journal of chemical theory and computation} \textbf{2021}, \emph{17}, 2355--2363\relax
\mciteBstWouldAddEndPuncttrue
\mciteSetBstMidEndSepPunct{\mcitedefaultmidpunct}
{\mcitedefaultendpunct}{\mcitedefaultseppunct}\relax
\EndOfBibitem
\bibitem[Pelaez \latin{et~al.}(2024)Pelaez, Simeon, Galvelis, Mirarchi, Eastman, Doerr, Th{\"o}lke, Markland, and De~Fabritiis]{pelaez2024torchmd}
Pelaez,~R.~P.; Simeon,~G.; Galvelis,~R.; Mirarchi,~A.; Eastman,~P.; Doerr,~S.; Th{\"o}lke,~P.; Markland,~T.~E.; De~Fabritiis,~G. TorchMD-Net 2.0: Fast Neural Network Potentials for Molecular Simulations. \emph{Journal of Chemical Theory and Computation} \textbf{2024}, \relax
\mciteBstWouldAddEndPunctfalse
\mciteSetBstMidEndSepPunct{\mcitedefaultmidpunct}
{}{\mcitedefaultseppunct}\relax
\EndOfBibitem
\bibitem[Acellera(2024)]{aceff_huggingface}
Acellera AceForce-1.0 on HuggingFace. \url{https://huggingface.co/Acellera/AceFF-1.0}, 2024; Accessed: 2025-02-24\relax
\mciteBstWouldAddEndPuncttrue
\mciteSetBstMidEndSepPunct{\mcitedefaultmidpunct}
{\mcitedefaultendpunct}{\mcitedefaultseppunct}\relax
\EndOfBibitem
\bibitem[De~Fabritiis(2024)]{nnpreviewgianni}
De~Fabritiis,~G. Machine Learning Potentials: A Roadmap Toward Next-Generation Biomolecular Simulations. \emph{arXiv preprint arXiv:2408.12625} \textbf{2024}, \relax
\mciteBstWouldAddEndPunctfalse
\mciteSetBstMidEndSepPunct{\mcitedefaultmidpunct}
{}{\mcitedefaultseppunct}\relax
\EndOfBibitem
\bibitem[Karwounopoulos \latin{et~al.}(2025)Karwounopoulos, Bieniek, Wu, Baskerville, Konig, Cossins, and Wood]{karwounopoulos2025evaluation}
Karwounopoulos,~J.; Bieniek,~M.; Wu,~Z.; Baskerville,~A.~L.; Konig,~G.; Cossins,~B.~P.; Wood,~G.~P. Evaluation of Machine Learning/Molecular Mechanics End-State Corrections with Mechanical Embedding to Calculate Relative Protein--Ligand Binding Free Energies. \emph{Journal of Chemical Theory and Computation} \textbf{2025}, \relax
\mciteBstWouldAddEndPunctfalse
\mciteSetBstMidEndSepPunct{\mcitedefaultmidpunct}
{}{\mcitedefaultseppunct}\relax
\EndOfBibitem
\bibitem[Sabane\'es~Zariquiey \latin{et~al.}(2024)Sabane\'es~Zariquiey, Galvelis, Gallicchio, Chodera, Markland, and De~Fabritiis]{sabanes2024enhancing}
Sabane\'es~Zariquiey,~F.; Galvelis,~R.; Gallicchio,~E.; Chodera,~J.~D.; Markland,~T.~E.; De~Fabritiis,~G. Enhancing Protein--Ligand Binding Affinity Predictions Using Neural Network Potentials. \emph{Journal of Chemical Information and Modeling} \textbf{2024}, \emph{64}, 1481--1485\relax
\mciteBstWouldAddEndPuncttrue
\mciteSetBstMidEndSepPunct{\mcitedefaultmidpunct}
{\mcitedefaultendpunct}{\mcitedefaultseppunct}\relax
\EndOfBibitem
\bibitem[Zeng \latin{et~al.}(2023)Zeng, Tao, Giese, and York]{zeng2023qdpi}
Zeng,~J.; Tao,~Y.; Giese,~T.~J.; York,~D.~M. QD$\pi$: A quantum deep potential interaction model for drug discovery. \emph{Journal of chemical theory and computation} \textbf{2023}, \emph{19}, 1261--1275\relax
\mciteBstWouldAddEndPuncttrue
\mciteSetBstMidEndSepPunct{\mcitedefaultmidpunct}
{\mcitedefaultendpunct}{\mcitedefaultseppunct}\relax
\EndOfBibitem
\bibitem[Tao \latin{et~al.}(2024)Tao, Giese, Ekesan, Zeng, Aradi, Hourahine, Aktulga, G{\"o}tz, Merz, and York]{tao2024amber}
Tao,~Y.; Giese,~T.~J.; Ekesan,~{\c{S}}.; Zeng,~J.; Aradi,~B.; Hourahine,~B.; Aktulga,~H.~M.; G{\"o}tz,~A.~W.; Merz,~K.~M.; York,~D.~M. Amber free energy tools: Interoperable software for free energy simulations using generalized quantum mechanical/molecular mechanical and machine learning potentials. \emph{The Journal of Chemical Physics} \textbf{2024}, \emph{160}\relax
\mciteBstWouldAddEndPuncttrue
\mciteSetBstMidEndSepPunct{\mcitedefaultmidpunct}
{\mcitedefaultendpunct}{\mcitedefaultseppunct}\relax
\EndOfBibitem
\bibitem[Crha \latin{et~al.}(2025)Crha, Poliak, Gillhofer, and Oostenbrink]{crha2025alchemical}
Crha,~R.; Poliak,~P.; Gillhofer,~M.; Oostenbrink,~C. Alchemical Free-Energy Calculations at Quantum-Chemical Precision. \emph{The Journal of Physical Chemistry Letters} \textbf{2025}, \emph{16}, 863--869\relax
\mciteBstWouldAddEndPuncttrue
\mciteSetBstMidEndSepPunct{\mcitedefaultmidpunct}
{\mcitedefaultendpunct}{\mcitedefaultseppunct}\relax
\EndOfBibitem
\bibitem[Galvelis \latin{et~al.}(2023)Galvelis, Varela-Rial, Doerr, Fino, Eastman, Markland, Chodera, and De~Fabritiis]{galvelis2023nnp}
Galvelis,~R.; Varela-Rial,~A.; Doerr,~S.; Fino,~R.; Eastman,~P.; Markland,~T.~E.; Chodera,~J.~D.; De~Fabritiis,~G. NNP/MM: Accelerating molecular dynamics simulations with machine learning potentials and molecular mechanics. \emph{Journal of chemical information and modeling} \textbf{2023}, \emph{63}, 5701--5708\relax
\mciteBstWouldAddEndPuncttrue
\mciteSetBstMidEndSepPunct{\mcitedefaultmidpunct}
{\mcitedefaultendpunct}{\mcitedefaultseppunct}\relax
\EndOfBibitem
\bibitem[Zeng \latin{et~al.}(2021)Zeng, Giese, Ekesan, and York]{zeng2021development}
Zeng,~J.; Giese,~T.~J.; Ekesan,~S.; York,~D.~M. Development of range-corrected deep learning potentials for fast, accurate quantum mechanical/molecular mechanical simulations of chemical reactions in solution. \emph{Journal of chemical theory and computation} \textbf{2021}, \emph{17}, 6993--7009\relax
\mciteBstWouldAddEndPuncttrue
\mciteSetBstMidEndSepPunct{\mcitedefaultmidpunct}
{\mcitedefaultendpunct}{\mcitedefaultseppunct}\relax
\EndOfBibitem
\bibitem[Azimi \latin{et~al.}(2022)Azimi, Khuttan, Wu, Pal, and Gallicchio]{azimi2022relative}
Azimi,~S.; Khuttan,~S.; Wu,~J.~Z.; Pal,~R.~K.; Gallicchio,~E. Relative binding free energy calculations for ligands with diverse scaffolds with the alchemical transfer method. \emph{J. Chem. Inf. Model.} \textbf{2022}, \emph{62}, 309--323\relax
\mciteBstWouldAddEndPuncttrue
\mciteSetBstMidEndSepPunct{\mcitedefaultmidpunct}
{\mcitedefaultendpunct}{\mcitedefaultseppunct}\relax
\EndOfBibitem
\bibitem[Saban\'es~Zariquiey \latin{et~al.}(2023)Saban\'es~Zariquiey, P\'esrez, Majewski, Gallicchio, and De~Fabritiis]{SabanesATM}
Saban\'es~Zariquiey,~F.; P\'esrez,~A.; Majewski,~M.; Gallicchio,~E.; De~Fabritiis,~G. Validation of the Alchemical Transfer Method for the Estimation of Relative Binding Affinities of Molecular Series. \emph{Journal of Chemical Information and Modeling} \textbf{2023}, \emph{63}, 2438--2444, PMID: 37042797\relax
\mciteBstWouldAddEndPuncttrue
\mciteSetBstMidEndSepPunct{\mcitedefaultmidpunct}
{\mcitedefaultendpunct}{\mcitedefaultseppunct}\relax
\EndOfBibitem
\bibitem[Chen \latin{et~al.}(2023)Chen, Wu, Wu, Silveira, Sherman, Xu, and Gallicchio]{ATMPsivant}
Chen,~L.; Wu,~Y.; Wu,~C.; Silveira,~A.; Sherman,~W.; Xu,~H.; Gallicchio,~E. Performance and Analysis of the Alchemical Transfer Method for Binding Free Energy Predictions of Diverse Ligands. 2023\relax
\mciteBstWouldAddEndPuncttrue
\mciteSetBstMidEndSepPunct{\mcitedefaultmidpunct}
{\mcitedefaultendpunct}{\mcitedefaultseppunct}\relax
\EndOfBibitem
\bibitem[Wang \latin{et~al.}(2015)Wang, Wu, Deng, Kim, Pierce, Krilov, Lupyan, Robinson, Dahlgren, Greenwood, Romero, Masse, Knight, Steinbrecher, Beuming, Damm, Harder, Sherman, Brewer, Wester, Murcko, Frye, Farid, Lin, Mobley, Jorgensen, Berne, Friesner, and Abel]{wang2015accurate}
Wang,~L.; Wu,~Y.; Deng,~Y.; Kim,~B.; Pierce,~L.; Krilov,~G.; Lupyan,~D.; Robinson,~S.; Dahlgren,~M.~K.; Greenwood,~J.; Romero,~D.~L.; Masse,~C.; Knight,~J.~L.; Steinbrecher,~T.; Beuming,~T.; Damm,~W.; Harder,~E.; Sherman,~W.; Brewer,~M.; Wester,~R.; Murcko,~M.; Frye,~L.; Farid,~R.; Lin,~T.; Mobley,~D.~L.; Jorgensen,~W.~L.; Berne,~B.~J.; Friesner,~R.~A.; Abel,~R. Accurate and reliable prediction of relative ligand binding potency in prospective drug discovery by way of a modern free-energy calculation protocol and force field. \emph{J. Am. Chem. Soc.} \textbf{2015}, \emph{137}, 2695--2703\relax
\mciteBstWouldAddEndPuncttrue
\mciteSetBstMidEndSepPunct{\mcitedefaultmidpunct}
{\mcitedefaultendpunct}{\mcitedefaultseppunct}\relax
\EndOfBibitem
\bibitem[Lu \latin{et~al.}(2021)Lu, Wu, Ghoreishi, Chen, Wang, Damm, Ross, Dahlgren, Russell, Von~Bargen, \latin{et~al.} others]{lu2021opls4}
Lu,~C.; Wu,~C.; Ghoreishi,~D.; Chen,~W.; Wang,~L.; Damm,~W.; Ross,~G.~A.; Dahlgren,~M.~K.; Russell,~E.; Von~Bargen,~C.~D.; others OPLS4: Improving force field accuracy on challenging regimes of chemical space. \emph{Journal of chemical theory and computation} \textbf{2021}, \emph{17}, 4291--4300\relax
\mciteBstWouldAddEndPuncttrue
\mciteSetBstMidEndSepPunct{\mcitedefaultmidpunct}
{\mcitedefaultendpunct}{\mcitedefaultseppunct}\relax
\EndOfBibitem
\bibitem[Acellera(2024)]{acellera_website}
Acellera Acellera Homepage. \url{https://www.acellera.com/}, 2024; Accessed: 2024-12-30\relax
\mciteBstWouldAddEndPuncttrue
\mciteSetBstMidEndSepPunct{\mcitedefaultmidpunct}
{\mcitedefaultendpunct}{\mcitedefaultseppunct}\relax
\EndOfBibitem
\bibitem[Sellers \latin{et~al.}(2017)Sellers, James, and Gobbi]{sellers2017comparison}
Sellers,~B.~D.; James,~N.~C.; Gobbi,~A. A comparison of quantum and molecular mechanical methods to estimate strain energy in druglike fragments. \emph{Journal of chemical information and modeling} \textbf{2017}, \emph{57}, 1265--1275\relax
\mciteBstWouldAddEndPuncttrue
\mciteSetBstMidEndSepPunct{\mcitedefaultmidpunct}
{\mcitedefaultendpunct}{\mcitedefaultseppunct}\relax
\EndOfBibitem
\bibitem[Gao \latin{et~al.}(2020)Gao, Ramezanghorbani, Isayev, Smith, and Roitberg]{gao2020torchani}
Gao,~X.; Ramezanghorbani,~F.; Isayev,~O.; Smith,~J.~S.; Roitberg,~A.~E. TorchANI: a free and open source PyTorch-based deep learning implementation of the ANI neural network potentials. \emph{Journal of chemical information and modeling} \textbf{2020}, \emph{60}, 3408--3415\relax
\mciteBstWouldAddEndPuncttrue
\mciteSetBstMidEndSepPunct{\mcitedefaultmidpunct}
{\mcitedefaultendpunct}{\mcitedefaultseppunct}\relax
\EndOfBibitem
\bibitem[Bannwarth \latin{et~al.}(2019)Bannwarth, Ehlert, and Grimme]{bannwarth2019gfn2}
Bannwarth,~C.; Ehlert,~S.; Grimme,~S. GFN2-xTB—An accurate and broadly parametrized self-consistent tight-binding quantum chemical method with multipole electrostatics and density-dependent dispersion contributions. \emph{Journal of chemical theory and computation} \textbf{2019}, \emph{15}, 1652--1671\relax
\mciteBstWouldAddEndPuncttrue
\mciteSetBstMidEndSepPunct{\mcitedefaultmidpunct}
{\mcitedefaultendpunct}{\mcitedefaultseppunct}\relax
\EndOfBibitem
\bibitem[Wang and Song(2016)Wang, and Song]{wang2016geometry}
Wang,~L.-P.; Song,~C. Geometry optimization made simple with translation and rotation coordinates. \emph{The Journal of chemical physics} \textbf{2016}, \emph{144}\relax
\mciteBstWouldAddEndPuncttrue
\mciteSetBstMidEndSepPunct{\mcitedefaultmidpunct}
{\mcitedefaultendpunct}{\mcitedefaultseppunct}\relax
\EndOfBibitem
\bibitem[SimonBoothroyd \latin{et~al.}(2023)SimonBoothroyd, Thompson, Mitchell, Wang, Wagner, and trevorgokey]{openff_recharge}
SimonBoothroyd; Thompson,~M.; Mitchell,~J.~A.; Wang,~L.; Wagner,~J.; trevorgokey openforcefield/openff-recharge: 0.5.0. 2023; \url{https://doi.org/10.5281/zenodo.8118623}\relax
\mciteBstWouldAddEndPuncttrue
\mciteSetBstMidEndSepPunct{\mcitedefaultmidpunct}
{\mcitedefaultendpunct}{\mcitedefaultseppunct}\relax
\EndOfBibitem
\bibitem[Galvelis \latin{et~al.}(2019)Galvelis, Doerr, Damas, Harvey, and De~Fabritiis]{galvelis2019scalable}
Galvelis,~R.; Doerr,~S.; Damas,~J.~M.; Harvey,~M.~J.; De~Fabritiis,~G. A scalable molecular force field parameterization method based on density functional theory and quantum-level machine learning. \emph{Journal of chemical information and modeling} \textbf{2019}, \emph{59}, 3485--3493\relax
\mciteBstWouldAddEndPuncttrue
\mciteSetBstMidEndSepPunct{\mcitedefaultmidpunct}
{\mcitedefaultendpunct}{\mcitedefaultseppunct}\relax
\EndOfBibitem
\bibitem[Doerr \latin{et~al.}(2016)Doerr, Harvey, No{\'e}, and De~Fabritiis]{doerr2016htmd}
Doerr,~S.; Harvey,~M.; No{\'e},~F.; De~Fabritiis,~G. HTMD: high-throughput molecular dynamics for molecular discovery. \emph{Journal of chemical theory and computation} \textbf{2016}, \emph{12}, 1845--1852\relax
\mciteBstWouldAddEndPuncttrue
\mciteSetBstMidEndSepPunct{\mcitedefaultmidpunct}
{\mcitedefaultendpunct}{\mcitedefaultseppunct}\relax
\EndOfBibitem
\bibitem[Gapsys \latin{et~al.}(2015)Gapsys, Michielssens, Peters, de~Groot, and Leonov]{gapsys2015calculation}
Gapsys,~V.; Michielssens,~S.; Peters,~J.~H.; de~Groot,~B.~L.; Leonov,~H. Calculation of binding free energies. \emph{Molecular Modeling of Proteins} \textbf{2015}, 173--209\relax
\mciteBstWouldAddEndPuncttrue
\mciteSetBstMidEndSepPunct{\mcitedefaultmidpunct}
{\mcitedefaultendpunct}{\mcitedefaultseppunct}\relax
\EndOfBibitem
\bibitem[Hahn \latin{et~al.}(2024)Hahn, Gapsys, de~Groot, Mobley, and Tresadern]{hahn2024current}
Hahn,~D.~F.; Gapsys,~V.; de~Groot,~B.~L.; Mobley,~D.~L.; Tresadern,~G. Current State of Open Source Force Fields in Protein--Ligand Binding Affinity Predictions. \emph{Journal of Chemical Information and Modeling} \textbf{2024}, \emph{64}, 5063--5076\relax
\mciteBstWouldAddEndPuncttrue
\mciteSetBstMidEndSepPunct{\mcitedefaultmidpunct}
{\mcitedefaultendpunct}{\mcitedefaultseppunct}\relax
\EndOfBibitem
\bibitem[Zou \latin{et~al.}(2019)Zou, Tian, and Simmerling]{zou2019blinded}
Zou,~J.; Tian,~C.; Simmerling,~C. Blinded prediction of protein--ligand binding affinity using Amber thermodynamic integration for the 2018 D3R grand challenge 4. \emph{J. Comput.-Aided Mol. Des.} \textbf{2019}, \emph{33}, 1021--1029\relax
\mciteBstWouldAddEndPuncttrue
\mciteSetBstMidEndSepPunct{\mcitedefaultmidpunct}
{\mcitedefaultendpunct}{\mcitedefaultseppunct}\relax
\EndOfBibitem
\bibitem[Maier \latin{et~al.}(2015)Maier, Martinez, Kasavajhala, Wickstrom, Hauser, and Simmerling]{maier2015ff14sb}
Maier,~J.~A.; Martinez,~C.; Kasavajhala,~K.; Wickstrom,~L.; Hauser,~K.~E.; Simmerling,~C. ff14SB: improving the accuracy of protein side chain and backbone parameters from ff99SB. \emph{J. Chem. Theory Comput.} \textbf{2015}, \emph{11}, 3696--3713\relax
\mciteBstWouldAddEndPuncttrue
\mciteSetBstMidEndSepPunct{\mcitedefaultmidpunct}
{\mcitedefaultendpunct}{\mcitedefaultseppunct}\relax
\EndOfBibitem
\bibitem[gpu()]{gpugrid}
GPUGRID - Volunteer Computing for Molecular Simulations. \url{https://www.gpugrid.net/}, Accessed: 2024-12-10\relax
\mciteBstWouldAddEndPuncttrue
\mciteSetBstMidEndSepPunct{\mcitedefaultmidpunct}
{\mcitedefaultendpunct}{\mcitedefaultseppunct}\relax
\EndOfBibitem
\bibitem[Eastman \latin{et~al.}(2023)Eastman, Galvelis, Pel{\'a}ez, Abreu, Farr, Gallicchio, Gorenko, Henry, Hu, Huang, \latin{et~al.} others]{eastman2023openmm}
Eastman,~P.; Galvelis,~R.; Pel{\'a}ez,~R.~P.; Abreu,~C.~R.; Farr,~S.~E.; Gallicchio,~E.; Gorenko,~A.; Henry,~M.~M.; Hu,~F.; Huang,~J.; others OpenMM 8: molecular dynamics simulation with machine learning potentials. \emph{The Journal of Physical Chemistry B} \textbf{2023}, \emph{128}, 109--116\relax
\mciteBstWouldAddEndPuncttrue
\mciteSetBstMidEndSepPunct{\mcitedefaultmidpunct}
{\mcitedefaultendpunct}{\mcitedefaultseppunct}\relax
\EndOfBibitem
\bibitem[Tan \latin{et~al.}(2012)Tan, Gallicchio, Lapelosa, and Levy]{tan2012theory}
Tan,~Z.; Gallicchio,~E.; Lapelosa,~M.; Levy,~R.~M. Theory of binless multi-state free energy estimation with applications to protein-ligand binding. \emph{The Journal of Chemical Physics} \textbf{2012}, \emph{136}, 04B608\relax
\mciteBstWouldAddEndPuncttrue
\mciteSetBstMidEndSepPunct{\mcitedefaultmidpunct}
{\mcitedefaultendpunct}{\mcitedefaultseppunct}\relax
\EndOfBibitem
\bibitem[Macdonald \latin{et~al.}(2022)Macdonald, dfhahn, Henry, Chodera, Dotson, Glass, and Pulido]{cinnabar}
Macdonald,~H.~B.; dfhahn; Henry,~M.; Chodera,~J.; Dotson,~D.; Glass,~W.; Pulido,~I. openforcefield/openff-arsenic: v0.2.1. 2022; \url{https://doi.org/10.5281/zenodo.6210305}\relax
\mciteBstWouldAddEndPuncttrue
\mciteSetBstMidEndSepPunct{\mcitedefaultmidpunct}
{\mcitedefaultendpunct}{\mcitedefaultseppunct}\relax
\EndOfBibitem
\bibitem[Gapsys \latin{et~al.}(2020)Gapsys, P{\'e}rez-Benito, Aldeghi, Seeliger, van Vlijmen, Gary, and de~Groot]{gapsys2020large}
Gapsys,~V.; P{\'e}rez-Benito,~L.; Aldeghi,~M.; Seeliger,~D.; van Vlijmen,~H.; Gary,~T.; de~Groot,~B. Large scale relative protein ligand binding affinities using non-equilibrium alchemy. 2020\relax
\mciteBstWouldAddEndPuncttrue
\mciteSetBstMidEndSepPunct{\mcitedefaultmidpunct}
{\mcitedefaultendpunct}{\mcitedefaultseppunct}\relax
\EndOfBibitem
\end{mcitethebibliography}

%\newpage
%\begin{figure}[h!]
%\centering
%\includegraphics[width=\columnwidth]%{Figures/RBFE_NNP_TOC.pdf}
%\caption*{For Table of Contents only}
%\label{fig:TOC}
%\end{figure}

\section{Supporting Information} \label{sec:sup_methods}
\setcounter{figure}{0}  
\setcounter{table}{0}  
\renewcommand{\thefigure}{S\arabic{figure}}
\renewcommand{\thetable}{S\arabic{table}}  

\begin{table}
\footnotesize
\centering
\begin{tabular}{c|c|c|c|c|c}
\hline
             $\lambda$   &  $\lambda_1$  & $\lambda_2$  & $\alpha$  &  $u_0$  &  $w_0$\\

\hline
 0.00  & 0.00 & 0.00 & 0.10 & 110 & 0 \\
 0.05  & 0.00 & 0.10 & 0.10 & 110 & 0 \\
 0.10  & 0.00 & 0.20 & 0.10 & 110 & 0 \\
 0.15  & 0.00 & 0.30 & 0.10 & 110 & 0 \\
 0.20  & 0.00 & 0.40 & 0.10 & 110 & 0 \\
 0.25  & 0.00 & 0.50 & 0.10 & 110 & 0 \\
 0.30  & 0.10 & 0.50 & 0.10 & 110 & 0 \\
 0.35  & 0.20 & 0.50 & 0.10 & 110 & 0 \\
 0.40  & 0.30 & 0.50 & 0.10 & 110 & 0 \\
 0.45  & 0.40 & 0.50 & 0.10 & 110 & 0 \\
 0.50  & 0.50 & 0.50 & 0.10 & 110 & 0 \\
 0.55  & 0.50 & 0.50 & 0.10 & 110 & 0 \\
 0.60  & 0.40 & 0.50 & 0.10 & 110 & 0 \\
 0.65  & 0.30 & 0.50 & 0.10 & 110 & 0 \\
 0.70  & 0.20 & 0.50 & 0.10 & 110 & 0 \\
 0.75  & 0.10 & 0.50 & 0.10 & 110 & 0 \\
 0.80  & 0.00 & 0.40 & 0.10 & 110 & 0 \\
 0.85  & 0.00 & 0.30 & 0.10 & 110 & 0 \\
 0.90  & 0.00 & 0.20 & 0.10 & 110 & 0 \\
 0.95  & 0.00 & 0.10 & 0.10 & 110 & 0 \\
 1.00  & 0.00 & 0.00 & 0.10 & 110 & 0 \\

\end{tabular}
\caption{\label{tab:params_table}Alchemical schedule of the Solftplus Alchemical Potential for the two legs for the alchemical transformations. $\alpha$ values are in (kcal/mol)$^{-1}$ and $u_0$ and $w_0$ are in kcal/mol }
\end{table}
\break
\subsection{\texorpdfstring{$\Delta G$ values}{Delta G values}}
\FloatBarrier
\begin{table*}[ht]
\centering
\resizebox{\textwidth}{!}{%
\begin{tabular}{l|ccc|ccc|ccc|c}
\textbf{Target} & \multicolumn{3}{c|}{\textbf{GAFF2}} & \multicolumn{3}{c|}{\textbf{OPLS4 (FEP+)\cite{lu2021opls4}}} & \multicolumn{3}{c|}{\textbf{AceFF 1.0}} & \textbf{{$N_{\text{lig}}$}} \\ 
 & RMSE & MAE & Kendall ($\tau$) & RMSE & MAE & Kendall ($\tau$) & RMSE & MAE & Kendall ($\tau$) & \\ 
\hline
BACE & 1.07\scriptsize$_{0.85}^{1.27}$ & 0.87\scriptsize$_{0.65}^{1.08}$ & 0.33\scriptsize$_{0.06}^{0.56}$ & \textbf{0.87\scriptsize$_{0.67}^{1.08}$} & \textbf{0.71\scriptsize$_{0.55}^{0.88}$} & 0.46\scriptsize$_{0.23}^{0.65}$ & 0.99\scriptsize$_{0.75}^{1.24}$ & 0.80\scriptsize$_{0.61}^{1.00}$ & \textbf{0.48\scriptsize$_{0.26}^{0.67}$} & 36 \\ 
CDK2 & 1.03\scriptsize$_{0.62}^{1.37}$ & 0.79\scriptsize$_{0.51}^{1.11}$ & 0.45\scriptsize$_{0.13}^{0.80}$ & \textbf{0.92\scriptsize$_{0.60}^{1.17}$} & 0.72\scriptsize$_{0.45}^{1.00}$ & 0.45\scriptsize$_{0.07}^{0.76}$ & 0.94\scriptsize$_{0.46}^{1.36}$ & \textbf{0.64\scriptsize$_{0.33}^{1.03}$} & \textbf{0.68\scriptsize$_{0.41}^{0.88}$} & 16 \\ 
JNK1 & 1.03\scriptsize$_{0.76}^{1.28}$ & 0.83\scriptsize$_{0.59}^{1.11}$ & 0.54\scriptsize$_{0.29}^{0.73}$ & \textbf{0.72\scriptsize$_{0.52}^{0.90}$} & \textbf{0.59\scriptsize$_{0.42}^{0.78}$} & \textbf{0.68\scriptsize$_{0.47}^{0.85}$} & 1.15\scriptsize$_{0.80}^{1.50}$ & 0.92\scriptsize$_{0.65}^{1.20}$ & 0.55\scriptsize$_{0.35}^{0.73}$ & 21 \\ 
MCL1 & 1.64\scriptsize$_{1.31}^{1.95}$ & 1.32\scriptsize$_{1.05}^{1.64}$ & 0.44\scriptsize$_{0.25}^{0.60}$ & \textbf{0.84\scriptsize$_{0.63}^{1.07}$} & \textbf{0.65\scriptsize$_{0.50}^{0.82}$} & \textbf{0.58\scriptsize$_{0.42}^{0.71}$} & 1.07\scriptsize$_{0.90}^{1.23}$ & 0.91\scriptsize$_{0.73}^{1.10}$ & 0.28\scriptsize$_{0.09}^{0.49}$ & 42 \\ 
P38 & 0.99\scriptsize$_{0.70}^{1.30}$ & 0.75\scriptsize$_{0.53}^{0.97}$ & 0.64\scriptsize$_{0.45}^{0.78}$ & \textbf{0.75\scriptsize$_{0.57}^{0.94}$} & \textbf{0.58\scriptsize$_{0.42}^{0.74}$} & 0.55\scriptsize$_{0.36}^{0.70}$ & 0.89\scriptsize$_{0.65}^{1.16}$ & 0.67\scriptsize$_{0.49}^{0.87}$ & \textbf{0.72\scriptsize$_{0.58}^{0.82}$} & 34 \\  
THROMBIN & 0.82\scriptsize$_{0.47}^{1.08}$ & 0.62\scriptsize$_{0.32}^{0.93}$ & 0.55\scriptsize$_{0.00}^{0.96}$ & \textbf{0.57\scriptsize$_{0.32}^{0.77}$} & \textbf{0.45\scriptsize$_{0.27}^{0.66}$} & \textbf{0.60\scriptsize$_{0.19}^{0.92}$} & 0.80\scriptsize$_{0.49}^{1.12}$ & 0.64\scriptsize$_{0.38}^{0.93}$ & 0.42\scriptsize$_{-0.15}^{0.84}$ & 11 \\ 
TYK2 & 0.65\scriptsize$_{0.44}^{0.82}$ & 0.52\scriptsize$_{0.35}^{0.71}$ & 0.67\scriptsize$_{0.42}^{0.84}$ & \textbf{0.45\scriptsize$_{0.24}^{0.63}$} & \textbf{0.34\scriptsize$_{0.21}^{0.50}$} & \textbf{0.80\scriptsize$_{0.56}^{0.96}$} & 0.74\scriptsize$_{0.53}^{0.92}$ & 0.59\scriptsize$_{0.37}^{0.83}$ & \textbf{0.80\scriptsize$_{0.56}^{0.96}$} & 16 \\ 
\hline
ALL & 1.17\scriptsize$_{1.03}^{1.30}$ & 0.90\scriptsize$_{0.79}^{1.01}$ & 0.55\scriptsize$_{0.48}^{0.62}$ & \textbf{0.78\scriptsize$_{0.69}^{0.88}$} & \textbf{0.61\scriptsize$_{0.53}^{0.68}$} & \textbf{0.66\scriptsize$_{0.60}^{0.72}$} & 0.99\scriptsize$_{0.89}^{1.10}$ & 0.79\scriptsize$_{0.71}^{0.89}$ & 0.59\scriptsize$_{0.52}^{0.65}$ & 176 \\ 
\end{tabular}%
}
\caption{\label{tab:stats_table} Comparison of the performance of different ligand force fields: GAFF2, OPLS4, and AceFF 1.0. Root Mean Square Error (RMSE) in kcal/mol and Kendall correlation ($\tau$) for the 8 tested protein targets. 95\% confidence intervals are shown with lower and upper bounds as subscripts and superscripts.}
\end{table*}
\FloatBarrier
\subsection{Evaluating the accuracy of calculations}
We analyzed the percentage of predictions that meet absolute error thresholds below 1 kcal/mol and 2 kcal/mol to evaluate how well AceFF 1.0 minimizes outliers in comparison to the other described methods. Table \ref{tab:MAE_pct_table} presents the percentage of values for each method across different protein targets within these thresholds, providing insight into the models' accuracy.
At the 2 kcal/mol threshold, AceFF 1.0 calculations demonstrate competitive accuracy across most targets, with fewer outliers indicated by higher percentages within this error range. For example, on targets like BACE and P38, AceFF 1.0 generally outperforms GAFF2, demonstrating AceFF’s 1.0 capacity to maintain accuracy across diverse systems. Similar trends are observed for THROMBIN and TYK2, where AceFF 1.0 shows strong performance relative to the other methods.
For predictions within at the 1 kcal/mol threshold, AceFF 1.0 remains competitive with OPLS4 and GAFF2. On CDK2, JNK1, and MCL1, AceFF 1.0 performs closely to OPLS4. Notably, on TYK2, AceFF achieves a higher percentage of values within the 1 kcal/mol range than either OPLS4 or GAFF2.
Overall, while OPLS4 reaches the highest absolute accuracy on some individual targets, AceFF 1.0 provides robust, consistent performance across both error thresholds, achieving comparable or superior accuracy across a range of protein targets.
\FloatBarrier
\begin{table*}[ht!]
\centering
\resizebox{\textwidth}{!}{%
\begin{tabular}{l|cc|cc|cc|c|c}
\textbf{Target} & \multicolumn{2}{c|}{\textbf{GAFF2}} & \multicolumn{2}{c|}{\textbf{OPLS4 (FEP+)}} & \multicolumn{2}{c|}{\textbf{AceFF 1.0}} \\
 & \textbf{\% MAE $<$ 1} & \textbf{\% MAE $<$ 2} & \textbf{\% MAE $<$ 1} & \textbf{\% MAE $<$ 2} & \textbf{\% MAE $<$ 1} & \textbf{\% MAE $<$ 2} \\
\hline
\textbf{BACE} & 60  & 78  & 59  & 93  & 66 & 93  \\
\textbf{CDK2} & 44  & 76  & 52  & 96  & 81  & 88  \\
\textbf{JNK1} & 65  & 91  & 71 & 100  & 56  & 82  \\
\textbf{MCL1} & 52 & 73  & 61  & 90  & 49  & 82 \\
\textbf{P38} & 71 & 95 & 66 & 100  & 63  & 93  \\
\textbf{PTP1B} & 59  & 84  & 94 & 98  & 57  & 82 \\
\textbf{THROMBIN} & 56  & 81  & 63 & 94 & 63  & 94 \\
\textbf{TYK2} & 78  & 96 & 79  & 96  & 92  & 100  \\
\end{tabular}%
}
\caption{\label{tab:MAE_pct_table} Percentage of predictions with a mean absolute error (MAE) below 1 kcal/mol and 2 kcal/mol for each target, comparing the performance of GAFF2, OPLS4 with FEP+, and AceFF 1.0, providing insight into each method\u2019s ability to minimize outliers and maintain accuracy across diverse protein systems.}
\end{table*}

\subsection{\texorpdfstring{$\Delta\Delta G$ values}{Delta Delta G values}}
\FloatBarrier
% DDG values here
% RMSE
\begin{table*}[ht]
\centering
\resizebox{\textwidth}{!}{%
\begin{tabular}{l|ccc|ccc|ccc|ccc|c}
\textbf{Target} & \multicolumn{3}{c|}{\textbf{GAFF2}} & \multicolumn{3}{c|}{\textbf{OPLS4 (FEP+)}} & \multicolumn{3}{c|}{\textbf{AceFF 1.0}} & \textbf{{$N_{\text{edge}}$}} \\
 & RMSE & MAE & Kendall ($\tau$) & RMSE & MAE & Kendall ($\tau$) & RMSE & MAE & Kendall ($\tau$) & \\
\hline
BACE & 1.48\scriptsize$_{1.20}^{1.75}$ & 1.16\scriptsize$_{0.92}^{1.39}$ & 0.29\scriptsize$_{0.10}^{0.48}$ & \textbf{1.02\scriptsize$_{0.85}^{1.19}$} & \textbf{0.82\scriptsize$_{0.67}^{0.98}$} & 0.37\scriptsize$_{0.20}^{0.51}$ & 1.32\scriptsize$_{1.04}^{1.61}$ & 1.04\scriptsize$_{0.84}^{1.27}$ & \textbf{0.46\scriptsize$_{0.31}^{0.58}$} & 58 \\
CDK2 & 1.50\scriptsize$_{1.14}^{1.83}$ & 1.21\scriptsize$_{0.87}^{1.58}$ & 0.20\scriptsize$_{-0.09}^{0.48}$ & \textbf{1.13\scriptsize$_{0.89}^{1.33}$} & 0.97\scriptsize$_{0.74}^{1.20}$ & 0.32\scriptsize$_{0.01}^{0.55}$ & 1.32\scriptsize$_{0.85}^{1.71}$ & \textbf{0.88\scriptsize$_{0.52}^{1.28}$} & \textbf{0.40\scriptsize$_{0.06}^{0.69}$} & 25 \\
JNK1 & 1.13\scriptsize$_{0.83}^{1.43}$ & 0.89\scriptsize$_{0.69}^{1.15}$ & 0.34\scriptsize$_{0.07}^{0.59}$ & \textbf{0.90\scriptsize$_{0.72}^{1.07}$} & \textbf{0.72\scriptsize$_{0.53}^{0.89}$} & \textbf{0.39\scriptsize$_{0.15}^{0.60}$} & 1.09\scriptsize$_{0.84}^{1.32}$ & 0.86\scriptsize$_{0.62}^{1.09}$ & 0.37\scriptsize$_{0.11}^{0.58}$ & 34 \\
MCL1 & 1.91\scriptsize$_{1.52}^{2.30}$ & 1.43\scriptsize$_{1.15}^{1.74}$ & \textbf{0.42\scriptsize$_{0.27}^{0.55}$} & \textbf{1.17\scriptsize$_{1.01}^{1.34}$} & 0.96\scriptsize$_{0.80}^{1.11}$ & 0.33\scriptsize$_{0.17}^{0.46}$ & 1.22\scriptsize$_{1.01}^{1.42}$ & \textbf{0.94\scriptsize$_{0.77}^{1.11}$} & 0.41\scriptsize$_{0.28}^{0.52}$ & 71 \\
P38 & 1.11\scriptsize$_{0.80}^{1.39}$ & 0.81\scriptsize$_{0.63}^{1.02}$ & 0.62\scriptsize$_{0.50}^{0.73}$ & \textbf{0.86\scriptsize$_{0.70}^{0.99}$} & \textbf{0.67\scriptsize$_{0.52}^{0.82}$} & 0.63\scriptsize$_{0.51}^{0.73}$ & 1.10\scriptsize$_{0.91}^{1.28}$ & 0.88\scriptsize$_{0.72}^{1.05}$ & \textbf{0.65\scriptsize$_{0.54}^{0.74}$} & 56 \\
THROMBIN & 1.42\scriptsize$_{0.97}^{1.85}$ & 1.12\scriptsize$_{0.72}^{1.55}$ & 0.41\scriptsize$_{0.12}^{0.69}$ & \textbf{1.02\scriptsize$_{0.72}^{1.31}$} & \textbf{0.86\scriptsize$_{0.61}^{1.15}$} & \textbf{0.41\scriptsize$_{0.03}^{0.68}$} & 1.36\scriptsize$_{0.75}^{1.88}$ & 1.00\scriptsize$_{0.57}^{1.47}$ & 0.18\scriptsize$_{-0.20}^{0.52}$ & 16 \\
TYK2 & 0.88\scriptsize$_{0.58}^{1.16}$ & 0.66\scriptsize$_{0.44}^{0.92}$ & 0.53\scriptsize$_{0.23}^{0.77}$ & 0.85\scriptsize$_{0.56}^{1.09}$ & 0.66\scriptsize$_{0.46}^{0.87}$ & \textbf{0.65\scriptsize$_{0.44}^{0.81}$} & \textbf{0.77\scriptsize$_{0.56}^{0.96}$} & \textbf{0.57\scriptsize$_{0.37}^{0.76}$} & \textbf{0.64\scriptsize$_{0.43}^{0.80}$} & 24 \\
\hline
ALL & 1.46\scriptsize$_{1.30}^{1.62}$ & 1.09\scriptsize$_{0.97}^{1.20}$ & 0.42\scriptsize$_{0.34}^{0.49}$ & \textbf{1.01\scriptsize$_{0.94}^{1.09}$} & \textbf{0.81\scriptsize$_{0.74}^{0.88}$} & 0.43\scriptsize$_{0.37}^{0.49}$ & 1.22\scriptsize$_{1.10}^{1.33}$ & 0.94\scriptsize$_{0.85}^{1.03}$ & \textbf{0.46\scriptsize$_{0.40}^{0.51}$} & 284 \\
\end{tabular}%
}
\caption{\label{tab:stats_ddg_table} Comparison of the performance of different ligand force fields: GAFF2, OPLS4 with FEP+\cite{lu2021opls4}, and AceFF 1.0. Root Mean Square Error (RMSE) in kcal/mol and Kendall correlation ($\tau$) for the 8 tested protein targets. 95\% confidence intervals are shown with lower and upper bounds as subscripts and superscripts.}
\end{table*}

\begin{figure*}
\centering
\includegraphics[width=\linewidth]{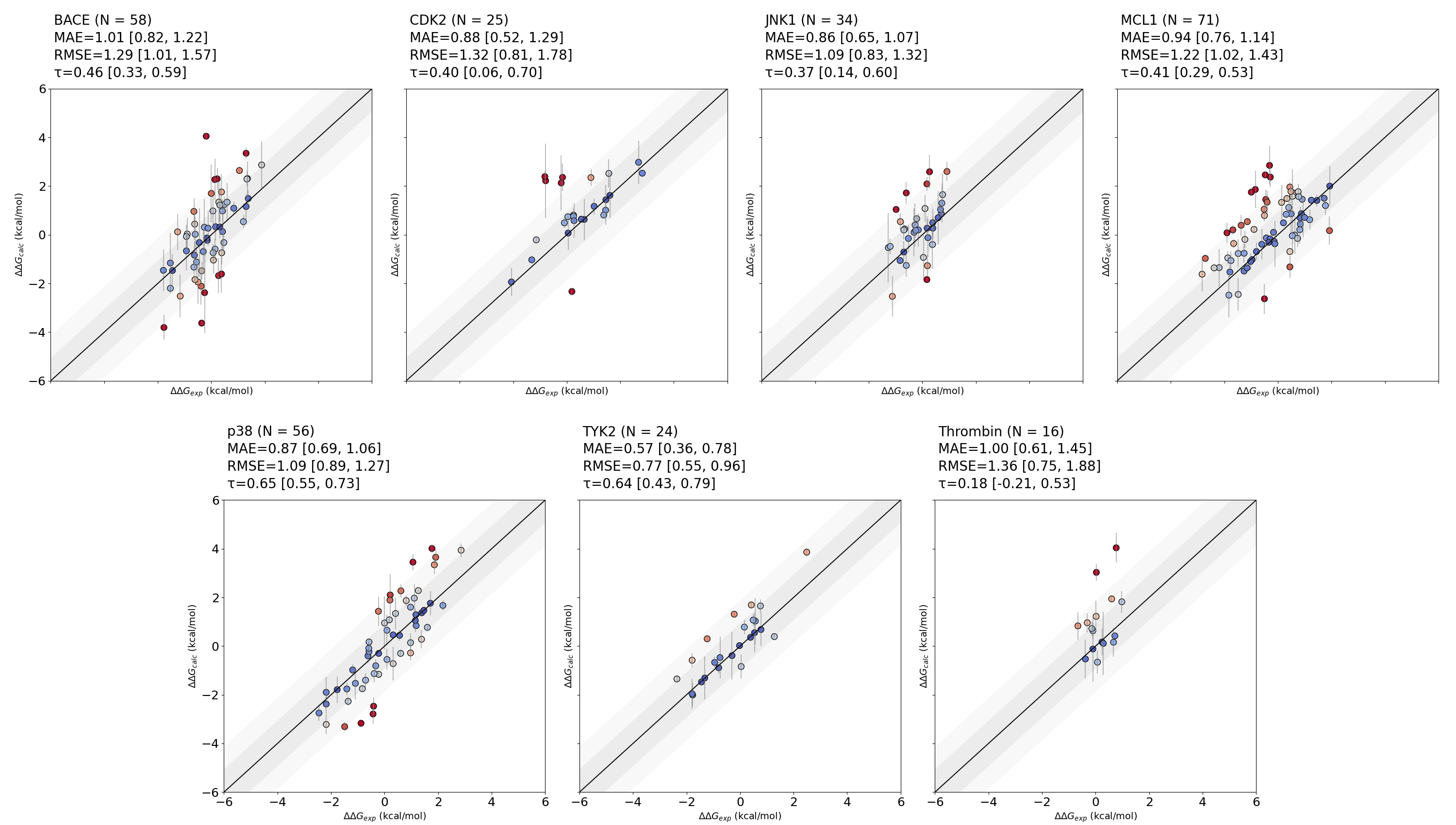}
\caption{Scatterplots of predicted $\Delta\Delta G$ values for each evaluated system using AceFF 1.0. The grey shaded areas represent absolute error thresholds of 1 kcal/mol and 2 kcal/mol. Additional metrics, including mean absolute error (MAE), root mean square error (RMSE) and Kendall tau correlation ($\tau$) are also displayed. 95\% confidence interval values are shown in brackets.}
\label{fig:scatter_ddG_AF}
\end{figure*}
\FloatBarrier
\subsection{RESP vs AM1BCC comparison}
\FloatBarrier
\begin{table*}[ht!]
\centering
\resizebox{\textwidth}{!}{%
\begin{tabular}{l|ccc|ccc|c}
\textbf{Target} & \multicolumn{3}{c|}{\textbf{RESP}} & \multicolumn{3}{c}{\textbf{AM1BCC}} &  \textbf{{$N_{\text{lig}}$}} \\ 
 & RMSE & MAE & Kendall ($\tau$) & RMSE & MAE & Kendall ($\tau$) & \\ 
\hline
BACE & 1.07\scriptsize$_{0.85}^{1.27}$ & 0.87\scriptsize$_{0.68}^{1.09}$ & 0.33\scriptsize$_{0.07}^{0.57}$ & 1.16\scriptsize$_{0.81}^{1.50}$ & 0.83\scriptsize$_{0.59}^{1.11}$ & 0.37\scriptsize$_{0.12}^{0.59}$ & 36 \\ 
CDK2 & 1.03\scriptsize$_{0.62}^{1.37}$ & 0.79\scriptsize$_{0.51}^{1.11}$ & 0.45\scriptsize$_{0.11}^{0.84}$ & 1.01\scriptsize$_{0.59}^{1.46}$ & 0.76\scriptsize$_{0.49}^{1.17}$ & 0.63\scriptsize$_{0.34}^{0.87}$ & 16 \\ 
JNK1 & 1.03\scriptsize$_{0.76}^{1.28}$ & 0.83\scriptsize$_{0.60}^{1.12}$ & 0.54\scriptsize$_{0.29}^{0.74}$ & 1.10\scriptsize$_{0.84}^{1.34}$ & 0.90\scriptsize$_{0.62}^{1.17}$ & 0.54\scriptsize$_{0.32}^{0.74}$ & 21 \\ 
MCL1 & 1.64\scriptsize$_{1.31}^{1.95}$ & 1.32\scriptsize$_{1.04}^{1.65}$ & 0.44\scriptsize$_{0.27}^{0.60}$ & 2.13\scriptsize$_{1.73}^{2.51}$ & 1.69\scriptsize$_{1.28}^{2.13}$ & 0.47\scriptsize$_{0.29}^{0.62}$ & 42 \\ 
P38 & 0.99\scriptsize$_{0.70}^{1.30}$ & 0.75\scriptsize$_{0.55}^{0.97}$ & 0.64\scriptsize$_{0.46}^{0.79}$ & 1.03\scriptsize$_{0.71}^{1.33}$ & 0.78\scriptsize$_{0.56}^{1.01}$ & 0.49\scriptsize$_{0.28}^{0.67}$ & 34 \\ 
PTP1B & 1.34\scriptsize$_{0.79}^{1.83}$ & 0.99\scriptsize$_{0.67}^{1.40}$ & 0.30\scriptsize$_{0.03}^{0.54}$ & 1.23\scriptsize$_{1.18}^{0.76}$ & 0.88\scriptsize$_{0.59}^{1.23}$ & 0.32\scriptsize$_{-0.04}^{0.61}$ & 23 \\ 
THROMBIN & 0.82\scriptsize$_{0.47}^{1.08}$ & 0.62\scriptsize$_{0.31}^{0.93}$ & 0.55\scriptsize$_{0.00}^{0.96}$ & 1.62\scriptsize$_{1.26}^{1.97}$ & 1.45\scriptsize$_{1.00}^{1.87}$ & 0.75\scriptsize$_{0.45}^{1.00}$ & 11 \\ 
TYK2 & 0.64\scriptsize$_{0.44}^{0.84}$ & 0.64\scriptsize$_{0.43}^{0.84}$ & 0.67\scriptsize$_{0.45}^{0.86}$ & 0.91\scriptsize$_{0.66}^{1.13}$ & 0.80\scriptsize$_{0.59}^{1.00}$ & 0.49\scriptsize$_{0.08}^{0.77}$ & 16 \\ 
\end{tabular}%
}
\caption{\label{tab:stats_charges_dG} Comparison charge models $\Delta G$}
\end{table*}
\FloatBarrier
%Top ids RESP vs AM1BCC
\begin{table*}[ht!]
\centering
\resizebox{\textwidth}{!}{%
\begin{tabular}{l|>{\centering\arraybackslash}p{2cm}>{\centering\arraybackslash}p{2cm}|>{\centering\arraybackslash}p{2cm}>{\centering\arraybackslash}p{2cm}|c}
\textbf{Target} & \multicolumn{2}{c|}{\textbf{RESP}} & \multicolumn{2}{c}{\textbf{AM1BCC}} & \textbf{{$N_{\text{lig}}$}} \\ 
 & \textbf{Top 5 (n/5)} & \textbf{FBL Top 30\% (\%)} & \textbf{Top 5 (n/5)} & \textbf{FBL Top 30\% (\%)} &  \\ 
\hline
BACE & 1 & 29 & 2 & 47 & 36 \\ 
CDK2 & 3 & 42 & 4 & 40 & 16 \\ 
JNK1 & 4 & 55 & 2 & 27 & 21 \\ 
MCL1 & 2 & 41 & 2 & 41 & 42 \\ 
P38 & 3 & 47 & 2 & 35 & 34 \\ 
PTP1B & 1 & 21 & 3 & 53 & 23 \\ 
THROMBIN & 3 & 72 & 3 & 67 & 11 \\ 
TYK2 & 4 & 29 & 3 & 21 & 16 \\ 
\end{tabular}%
}
\caption{\label{tab:topid_table_charges} Comparison of top identifiers between RESP and AM1BCC charge models.}
\end{table*}
\FloatBarrier
\begin{table*}[ht!]
\centering
\resizebox{\textwidth}{!}{%
\begin{tabular}{l|ccc|ccc|c}
\textbf{Target} & \multicolumn{3}{c|}{\textbf{RESP}} & \multicolumn{3}{c}{\textbf{AM1BCC}} &  \textbf{{$N_{\text{edges}}$}} \\ 
 & RMSE & MAE & Kendall ($\tau$) & RMSE & MAE & Kendall ($\tau$) & \\ 
\hline
BACE & 1.48\scriptsize$_{1.20}^{1.74}$ & 1.16\scriptsize$_{0.92}^{1.41}$ & 0.29\scriptsize$_{0.09}^{0.48}$ & 1.60\scriptsize$_{1.15}^{1.76}$ & 1.20\scriptsize$_{1.05}^{1.38}$ & 0.24\scriptsize$_{0.02}^{0.48}$ & 58 \\ 
CDK2 & 1.50\scriptsize$_{1.11}^{1.82}$ & 1.21\scriptsize$_{0.87}^{1.58}$ & 0.20\scriptsize$_{-0.07}^{0.47}$ & 1.26\scriptsize$_{0.95}^{1.56}$ & 1.00\scriptsize$_{0.69}^{1.27}$ & 0.35\scriptsize$_{0.14}^{0.57}$ & 25 \\ 
JNK1 & 1.13\scriptsize$_{0.82}^{1.42}$ & 0.89\scriptsize$_{0.67}^{1.12}$ & 0.34\scriptsize$_{0.06}^{0.60}$ & 1.07\scriptsize$_{0.74}^{1.34}$ & 0.93\scriptsize$_{0.72}^{1.15}$ & 0.33\scriptsize$_{0.08}^{0.64}$ & 34 \\ 
MCL1 & 1.91\scriptsize$_{1.52}^{2.33}$ & 1.43\scriptsize$_{1.13}^{1.73}$ & 0.42\scriptsize$_{0.26}^{0.56}$ & 2.61\scriptsize$_{2.15}^{2.91}$ & 2.00\scriptsize$_{1.68}^{2.28}$ & 0.27\scriptsize$_{0.12}^{0.42}$ & 71 \\ 
P38 & 1.11\scriptsize$_{0.80}^{1.40}$ & 0.81\scriptsize$_{0.63}^{1.02}$ & 0.62\scriptsize$_{0.50}^{0.73}$ & 1.47\scriptsize$_{1.14}^{1.73}$ & 1.10\scriptsize$_{0.90}^{1.21}$ & 0.51\scriptsize$_{0.32}^{0.67}$ & 56 \\ 
PTP1B & 1.98\scriptsize$_{1.24}^{2.64}$ & 1.33\scriptsize$_{0.94}^{1.83}$ & 0.17\scriptsize$_{-0.09}^{0.40}$ & 1.66\scriptsize$_{1.18}^{1.86}$ & 1.17\scriptsize$_{0.87}^{1.43}$ & 0.44\scriptsize$_{0.21}^{0.61}$ & 44 \\ 
THROMBIN & 1.42\scriptsize$_{0.95}^{1.86}$ & 1.12\scriptsize$_{0.74}^{1.54}$ & 0.41\scriptsize$_{0.10}^{0.67}$ & 1.90\scriptsize$_{1.46}^{2.32}$ & 1.54\scriptsize$_{1.10}^{1.97}$ & 0.38\scriptsize$_{0.15}^{0.65}$ & 16 \\ 
TYK2 & 0.88\scriptsize$_{0.60}^{1.15}$ & 0.66\scriptsize$_{0.43}^{0.89}$ & 0.53\scriptsize$_{0.24}^{0.74}$ & 1.32\scriptsize$_{1.10}^{1.58}$ & 1.10\scriptsize$_{0.90}^{1.19}$ & 0.37\scriptsize$_{0.28}^{0.57}$ & 24 \\ 
\end{tabular}%
}
\caption{\label{tab:stats_charges_ddG} Comparison charge models $\Delta\Delta G$}
\end{table*}
\FloatBarrier
\subsection{Top id comparison}
% Top id table
\FloatBarrier
\begin{table*}[ht!]
\centering
\resizebox{\textwidth}{!}{%
\begin{tabular}{l|cc|cc|cc|c}
\textbf{Target} & \multicolumn{2}{c|}{\textbf{GAFF2}} & \multicolumn{2}{c|}{\textbf{OPLS4 (FEP+)}} & \multicolumn{2}{c|}{\textbf{AceFF 1.0}} & \textbf{{$N_{\text{lig}}$}} \\
 & \textbf{Top 5} & \textbf{Top 30\%} & \textbf{Top 5} & \textbf{Top 30\%} & \textbf{Top 5} & \textbf{Top 30\%} & \\
 & \scriptsize(n/5) & \scriptsize(\%) & \scriptsize(n/5) & \scriptsize(\%) & \scriptsize(n/5) & \scriptsize(\%) & \\
\hline
BACE & 1 & 29 & \textbf{3} & \textbf{60} & \textbf{3} & \textbf{60} & 36 \\
CDK2 & \textbf{3} & 42 & \textbf{3} & 42 & 2 & \textbf{52} & 16 \\
JNK1 & \textbf{4} & \textbf{55} & \textbf{4} & \textbf{76} & 3 & 43 & 21 \\
MCL1 & 2 & 41 & \textbf{3} & \textbf{58} & 1 & 16 & 42 \\
P38 & 3 & 47 & 2 & 33 & \textbf{3} & \textbf{50} & 34 \\
PTP1B & 1 & 21 & \textbf{4} & \textbf{83} & 3 & 84 & 23 \\
THROMBIN & 3 & \textbf{72} & \textbf{4} & \textbf{72} & 2 & 28 & 11 \\
TYK2 & 4 & 29 & \textbf{5} & \textbf{73} & \textbf{5} & \textbf{73} & 16 \\
\end{tabular}%
}
\caption{\label{tab:topid_table} Comparison of top molecules identified (Top 5 and Top 30\%) across different methods.}
\end{table*}
\FloatBarrier
\break
\subsection{1fs vs 2fs}
\FloatBarrier
%% Comparison 1fs vs 2fs Tables

\begin{table*}[!ht]
\centering
\resizebox{\textwidth}{!}{%
\begin{tabular}{l|ccc|ccc|c}
\textbf{Target} & \multicolumn{3}{c|}{\textbf{timestep:1fs}} & \multicolumn{3}{c|}{\textbf{timestep:2fs}} &  \textbf{$N_{\text{lig}}$} \\ 
 & RMSE & MAE & Kendall ($\tau$) & RMSE & MAE & Kendall ($\tau$) & \\ 
\hline
BACE & 1.07\scriptsize$_{0.83}^{1.37}$ & 0.86\scriptsize$_{0.66}^{1.09}$ & \textbf{0.52\scriptsize$_{0.32}^{0.72}$} & \textbf{0.99\scriptsize$_{0.74}^{1.26}$ }& \textbf{0.79\scriptsize$_{0.61}^{1.00}$} & 0.48\scriptsize$_{0.27}^{0.67}$ & 36 \\ 
CDK2 & \textbf{0.90\scriptsize$_{0.42}^{1.32}$} & \textbf{0.68\scriptsize$_{0.37}^{1.00}$} & \textbf{0.65\scriptsize$_{0.40}^{0.89}$} & 0.94\scriptsize$_{0.46}^{1.36}$ & 0.64\scriptsize$_{0.33}^{1.03}$ & \textbf{0.68\scriptsize$_{0.41}^{0.88}$} & 16 \\ 
JNK1 & \textbf{1.12\scriptsize$_{0.91}^{1.32}$} & 0.98\scriptsize$_{0.76}^{1.22}$ & 0.50\scriptsize$_{0.28}^{0.72}$ & 1.15\scriptsize$_{0.80}^{1.50}$ & \textbf{0.92\scriptsize$_{0.66}^{1.21}$} & \textbf{0.55\scriptsize$_{0.35}^{0.73}$} & 21 \\ 
P38 & \textbf{0.77\scriptsize$_{0.55}^{0.97}$} & \textbf{0.58\scriptsize$_{0.41}^{0.76}$} & \textbf{0.72\scriptsize$_{0.60}^{0.82}$} & 0.89\scriptsize$_{0.62}^{1.13}$ & 0.67\scriptsize$_{0.49}^{0.87}$ & \textbf{0.72\scriptsize$_{0.60}^{0.82}$} & 34 \\ 
THROMBIN & \textbf{0.57\scriptsize$_{0.42}^{0.71}$} & \textbf{0.53\scriptsize$_{0.41}^{0.65}$} & \textbf{0.44\scriptsize$_{0.02}^{0.82}$} & 0.80\scriptsize$_{0.47}^{1.08}$ & 0.64\scriptsize$_{0.37}^{0.94}$ & 0.42\scriptsize$_{-0.14}^{0.83}$ & 11 \\ 
TYK2 & \textbf{0.47\scriptsize$_{0.23}^{0.66}$} &\textbf{ 0.31\scriptsize$_{0.16}^{0.51}$} & \textbf{0.85\scriptsize$_{0.66}^{0.98}$} & 0.74\scriptsize$_{0.53}^{0.92}$ & 0.59\scriptsize$_{0.38}^{0.81}$ & 0.80\scriptsize$_{0.58}^{0.96}$ & 16 \\ 
\end{tabular}%
}
\caption{\label{tab:stats_timestep_dG} Comparison of $\Delta G$ values from AceFF RBFE calculations at different timesteps}
\end{table*}
\FloatBarrier
\begin{table*}[!ht]
\centering
\resizebox{\textwidth}{!}{%
\begin{tabular}{l|ccc|ccc|c}
\textbf{Target} & \multicolumn{3}{c|}{\textbf{timestep:1fs}} & \multicolumn{3}{c|}{\textbf{timestep:2fs}} &  \textbf{$N_{\text{edges}}$} \\ 
 & RMSE & MAE & Kendall ($\tau$) & RMSE & MAE & Kendall ($\tau$) & \\ 
\hline
BACE & \textbf{1.20\scriptsize$_{0.92}^{1.47}$} & \textbf{0.89\scriptsize$_{0.69}^{1.10}$} & \textbf{0.49\scriptsize$_{0.35}^{0.62}$} & 1.32\scriptsize$_{1.05}^{1.62}$ & 1.04\scriptsize$_{0.83}^{1.27}$ & 0.46\scriptsize$_{0.31}^{0.59}$ & 58 \\ 
CDK2 & 1.35\scriptsize$_{0.87}^{1.25}$ & 0.82\scriptsize$_{0.50}^{1.24}$ & \textbf{0.42\scriptsize$_{0.10}^{0.72}$} & \textbf{1.32\scriptsize$_{0.85}^{1.17}$} & \textbf{0.88\scriptsize$_{0.52}^{1.28}$} & 0.40\scriptsize$_{0.06}^{0.70}$ & 25 \\ 
JNK1 & 1.22\scriptsize$_{0.95}^{1.42}$ & 0.98\scriptsize$_{0.74}^{1.22}$ & 0.22\scriptsize$_{-0.05}^{0.47}$ & \textbf{1.09\scriptsize$_{0.84}^{1.30}$} & \textbf{0.86\scriptsize$_{0.65}^{1.09}$} & \textbf{0.37\scriptsize$_{0.14}^{0.59}$} & 34 \\ 
P38 & \textbf{1.04\scriptsize$_{0.85}^{1.20}$} & \textbf{0.84\scriptsize$_{0.69}^{1.01}$} & \textbf{0.69\scriptsize$_{0.59}^{0.78}$} & 1.10\scriptsize$_{0.90}^{1.28}$ & 0.88\scriptsize$_{0.70}^{1.06}$ & 0.65\scriptsize$_{0.54}^{0.73}$ & 56 \\ 
THROMBIN & \textbf{1.07\scriptsize$_{0.74}^{1.36}$} & \textbf{0.89\scriptsize$_{0.63}^{1.20}$} &\textbf{ 0.28\scriptsize$_{-0.07}^{0.58}$} & 1.36\scriptsize$_{0.75}^{1.88}$ & 1.00\scriptsize$_{0.57}^{1.47}$ & 0.18\scriptsize$_{-0.20}^{0.52}$ & 16 \\ 
TYK2 & \textbf{0.55\scriptsize$_{0.36}^{0.73}$} & \textbf{0.42\scriptsize$_{0.29}^{0.58}$} & \textbf{0.73\scriptsize$_{0.53}^{0.88}$} & 0.77\scriptsize$_{0.56}^{0.96}$ & 0.57\scriptsize$_{0.37}^{0.76}$ & 0.64\scriptsize$_{0.43}^{0.80}$ & 24 \\ 
\end{tabular}%
}
\caption{\label{tab:stats_timestep_ddG} Comparison of $\Delta\Delta G$ values from AceFF RBFE calculations at different timesteps}
\end{table*}
\FloatBarrier
\begin{table*}[ht!]
\centering
\resizebox{\textwidth}{!}{%
\begin{tabular}{l|>{\centering\arraybackslash}p{2cm}>{\centering\arraybackslash}p{2cm}|>{\centering\arraybackslash}p{2cm}>{\centering\arraybackslash}p{2cm}|c}
\textbf{Target} & \multicolumn{2}{c|}{\textbf{timestep:1fs}} & \multicolumn{2}{c|}{\textbf{timestep:2fs}} & \textbf{$N_{\text{lig}}$} \\ 
 & \textbf{Top 5 (n/5)} & \textbf{FBL Top 30\% (\%)} & \textbf{Top 5 (n/5)} & \textbf{FBL Top 30\% (\%)} &  \\ 
\hline
BACE & \textbf{3} & \textbf{75} & \textbf{3} & 59 & 36 \\ 
CDK2 & \textbf{4} & \textbf{54} & 3 & 54 & 16 \\ 
JNK1 & 2 & 32 & \textbf{3} & \textbf{43} & 21 \\ 
P38 & 2 & 47 & \textbf{3} & \textbf{50} & 34 \\ 
THROMBIN & \textbf{3} & \textbf{56} & \textbf{3} & 11 & 11 \\ 
TYK2 & 4 & 81 & \textbf{5} & \textbf{85} & 16 \\ 
\end{tabular}%
}
\caption{\label{tab:topid_table_timestep} Comparison of top identifiers between AceFF 1fs and 2fs runs}
\end{table*}
\FloatBarrier
\break
\subsection{AceFF 1.0 vs ANI-2x}

\begin{table*}[ht]
\centering
\captionsetup{justification=centering}
\caption{Comparison of AceFF and ANI-2x calculations for $\Delta G$ values. Only the P38 and TYK2 targets are shown since these are the ones that we could compute all edges in our previous study with ANI-2x}
\label{tab:stats_ani}
\resizebox{\textwidth}{!}{%
\begin{tabular}{l|c|c|c|c|c|c|c}
%\hline

%\hline
\multirow{2}{*}{\textbf{Target}} & \multirow{2}{*}{$N_{\text{lig}}$} & \multicolumn{3}{c|}{\textbf{AceFF 1.0}} & \multicolumn{3}{c}{\textbf{ANI-2x}} \\
\cline{3-8}
 & & RMSE & MAE & Kendall ($\tau$) & RMSE & MAE & Kendall ($\tau$) \\
%\hline
P38 & 34 & \textbf{0.77\scriptsize$_{0.56}^{0.97}$} & \textbf{0.58\scriptsize$_{0.42}^{0.75}$} & \textbf{0.72\scriptsize$_{0.60}^{0.81}$} & 0.93\scriptsize$_{0.75}^{1.11}$ & 0.78\scriptsize$_{0.61}^{0.96}$ & 0.58\scriptsize$_{0.41}^{0.73}$ \\
TYK2 & 16 & \textbf{0.47\scriptsize$_{0.23}^{0.68}$} & \textbf{0.31\scriptsize$_{0.15}^{0.49}$} & \textbf{0.85\scriptsize$_{0.65}^{1.00}$} & 0.50\scriptsize$_{0.28}^{0.72}$ & 0.39\scriptsize$_{0.26}^{0.54}$ & 0.82\scriptsize$_{0.57}^{0.96}$ \\
%\hline
\end{tabular}%
}
\end{table*}
\FloatBarrier
\begin{table*}[ht]
\centering
\captionsetup{justification=centering}
\caption{Comparison of AceFF and ANI2x calculations for $\Delta \Delta G$ values. The CDK2 and JNK1 datasets are incomplete due to ANI-2x limitations.}
\label{tab:stats_ddg_ani}
\resizebox{\textwidth}{!}{%
\begin{tabular}{l|c|c|c|c|c|c|c}
\multirow{2}{*}{\textbf{Target}} & \multirow{2}{*}{$N_{\text{edges}}$} & \multicolumn{3}{c|}{\textbf{AceFF 1.0}} & \multicolumn{3}{c}{\textbf{ANI2x}} \\
\cline{3-8}
 & & \textbf{RMSE} & \textbf{MAE} & kendall ($\tau$) & \textbf{RMSE} & \textbf{MAE} & kendall ($\tau$) \\
%\hline
CDK2 & 22 & 1.03\scriptsize$_{0.57}^{1.41}$ & 0.67\scriptsize$_{0.35}^{1.03}$ & 0.46\scriptsize$_{0.34}^{0.59}$ & \textbf{0.83\scriptsize$_{0.62}^{1.03}$} & \textbf{0.72\scriptsize$_{0.58}^{0.94}$} & \textbf{0.62\scriptsize$_{0.34}^{0.81}$} \\
JNK1 & 27 & \textbf{0.92\scriptsize$_{0.70}^{1.13}$} & \textbf{0.74\scriptsize$_{0.56}^{0.94}$} & \textbf{0.46\scriptsize$_{0.21}^{0.67}$} & 0.90\scriptsize$_{0.74}^{1.06}$ & 0.68\scriptsize$_{0.51}^{0.80}$ & 0.43\scriptsize$_{0.30}^{0.55}$ \\
P38 & 56 & \textbf{1.04\scriptsize$_{0.85}^{1.19}$} & \textbf{0.84\scriptsize$_{0.69}^{1.00}$} & \textbf{0.69\scriptsize$_{0.59}^{0.78}$} & 1.17\scriptsize$_{0.94}^{1.37}$ & 0.91\scriptsize$_{0.72}^{1.10}$ & 0.59\scriptsize$_{0.48}^{0.70}$ \\
TYK2 & 24 & \textbf{0.55\scriptsize$_{0.36}^{0.72}$} & \textbf{0.42\scriptsize$_{0.28}^{0.57}$} & \textbf{0.73\scriptsize$_{0.54}^{0.88}$} & 0.56\scriptsize$_{0.43}^{0.68}$ & 0.47\scriptsize$_{0.36}^{0.59}$ & 0.67\scriptsize$_{0.42}^{0.82}$ \\
\end{tabular}%
}
\end{table*}
\break
\subsection{Quantumbind-RBFE speed}

\FloatBarrier
\begin{figure*}
\centering
\includegraphics[width=\linewidth]{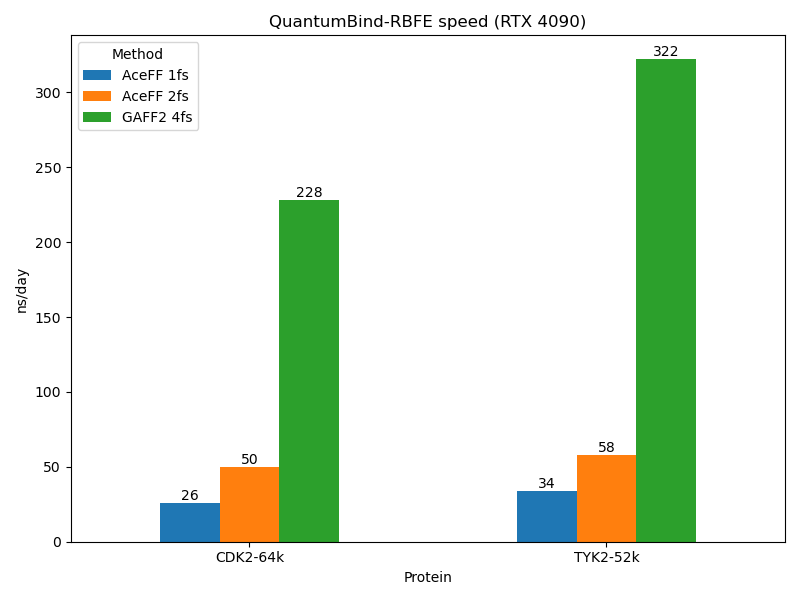}
\caption{QuantumBind-RBFE simulation speed for various modes. AceFF at 1 and 2fs and GAFF2 at 4fs. All evaluations are done with a RTX 4090 GPU}
\label{fig:QB_speeds}
\end{figure*}
\FloatBarrier
\FloatBarrier

\end{document}